\documentstyle[preprint,aps,epsf]{revtex}
\newif\iftightenlines\tightenlinesfalse
\tightenlines\tightenlinestrue

\def\eslt{\not\!\!{E_T}}
\def\to{\rightarrow}

\def\te{\tilde e}
\def\tl{\tilde l}

\def\ttau{\tilde \tau}
\def\tmu{\tilde \mu}

\def\tnu{\tilde\nu}
\def\tell{\tilde\ell}

\def\tw{\widetilde W}
\def\tz{\widetilde Z}
\begin{document}
\preprint{\vbox{\baselineskip=14pt%
   \rightline{UH-511-990-01}
   \rightline{FSU-HEP-010717}
}}
\title{Sneutrino Mass Measurements at $e^+e^-$ Linear Colliders}

\author{J.~Kenichi Mizukoshi$^1$, Howard Baer$^2$, A.~S.~Belyaev$^2$ and
Xerxes Tata$^1$}
\address{
$^1$Department of Physics and Astronomy,
University of Hawaii,
Honolulu, HI 96822, USA}
\address{
$^2$Department of Physics,
Florida State University,
Tallahassee, FL 32306 USA}
%
\maketitle
\begin{abstract}
It is generally accepted that experiments at an $e^+e^-$ linear collider
will be able to extract the masses of the selectron as well as the
associated sneutrino with a precision of $\sim 1$\% by determining the
kinematic end points of the energy spectrum of daughter electrons
produced in their two body decays to a lighter neutralino or
chargino. Recently, it has been suggested that by studying the energy
dependence of the cross section near the production threshold, this
precison can be improved by an order of magnitude, assuming an
integrated luminosity of 100~fb$^{-1}$. It is further suggested that
these threshold scans also allow the masses of even the heavier second
and third generation sleptons and sneutrinos to be determined to better
than 0.5\%.  We re-examine the prospects for determining sneutrino
masses. We find that the cross sections for the second and third
generation sneutrinos are too small for a threshold scan to be
useful. An additonal complication arises because the cross section for
sneutrino pairs to decay into any visible final state(s) necessarily
depends on an unknown branching fraction, so that the overall
normalization is unknown. This reduces the precision with which the
sneutrino mass can be extracted.  We propose a different strategy to
optimize the extraction of $m(\tnu_{\mu})$ and $m(\tnu_{\tau})$ via the
energy dependence of the cross section. We find that even with an
integrated luminosity of 500~fb$^{-1}$, these can be determined with a
precision no better than several percent at the 90\% CL. We also
examine the measurement of $m(\tnu_e)$ and show that it can be extracted
with a precision of about 0.5\% (0.2\%) with an integrated luminosity of
120~fb$^{-1}$ (500~fb$^{-1}$).

\end{abstract}

\medskip

%
\newpage

\section{Introduction}\label{sec1}

There have been many studies\cite{barklow,snow96} that have pointed out
the complementarity between experiments at the Large Hadron Collider
(LHC) and at $e^+e^-$ linear colliders (LC) that are being considered as
the next major high energy accelerator facility after the LHC. High
energy hadron colliders with general purpose hermetic detectors are
ideal for broad band searches of new phenomena\cite{atlas,cms}.  While
some recent studies (performed within the context of specific
supersymmetric models) have shown that it may be possible to make
precision measurements of masses (more specifically, mass
differences)\cite{atlas,lhcsnow,hinch,hinchgmsb} and possibly other
attributes of new particles \cite{ambrolhc}, experiments at the LC allow
a relatively model-independent determination of masses
\cite{jlc,mur,munroe,zdr,snownlc,tesla} and other properties of Higgs
bosons and supersymmetric (SUSY) particles assuming that these are
kinematically accessible, in addition to more
detailed\cite{ambronlc,mercadante} measurements specific to particular
models.

The original studies \cite{jlc,mur} of sparticle mass measurements at
linear colliders rely largely on kinematic reconstruction of masses;
{\it e.g.} for the measurement of $m(\te_R)$, the energy distributions
of the electrons in $\te_R \to e\tz_1$ is, except for effects of cuts
and resolution smearing, flat with sharp end points determined only by
$m(\te_R)$ and $m(\tz_1)$. It has been shown that with several tens of
fb$^{-1}$, experiments at LCs should measure these masses with a
precision at the percent level. A similar precision was shown to be
possible for the determination of the lighter chargino mass, even though
the energy spectrum of the visible daughters is not flat.  The same idea
was extended to sparticles decaying via cascades; {\it e.g}. $\tnu_e \to
e\tw_1$, $\tw_1 \to f\overline{f'}\tz_1$, ($f=q,\mu$). The end points of the
electron energy distribution are determined by just $m(\tnu_e)$ and
$m(\tw_1)$. Again, these sparticle masses were shown to be measureable
at the 1-2\% level \cite{munroe,zdr,snownlc}. In all these studies,
longitudinal polarization of the electron beam was essential to control
Standard Model (SM) backgrounds and SUSY contamination.

Very recently, in the TESLA Technical Design Report \cite{tesla}, it has
been emphasized that the tunability of the energy of a LC may be used to
perform an energy scan close to a new production threshold. It is
claimed that in some cases ($\tw_1, \te_L, \te_R, \tnu_e$) this allows
sparticle masses to be determined with a precision of a part per mille
at a LC. Specifically, it was suggested that this could be achieved by a
threshold scan of the cross section (10 points spaced 1 GeV apart with
an integrated luminosity of 10~fb$^{-1}$/point) in any particular
channel chosen so that SM backgrounds and SUSY contamination is
small. Further, these results were extrapolated to argue that masses of
the heavier sleptons of the second and third generations
$(\tmu_L,\tnu_{\mu},\ttau_2$ and $\tnu_{\tau})$ could also be measured
with a precision of about 0.5\%. For the measurements of third
generation sleptons, it was assumed that secondary vertex detection
would serve to efficiently tag tau leptons. Without making any
representation about whether or not this is possible, we will assume
this to be true for our analysis. We will, therefore,
optimistically assume that light flavour and gluon jets are not a
background for hadronically decaying taus, and further, that
leptonically decaying isolated taus will always be distinguished from
prompt $e$s and
$\mu$s by their displaced vertices.

If sparticle masses can indeed be determined with the impressive precision
listed in Ref.\cite{tesla} it should serve to strigently test various models of
how SUSY breaking is mediated to the superpartners of SM
particles. Blair, Porod and Zerwas \cite{porod} have clearly illustrated
how such measurements could be used to test scalar mass unification
expected within the mSUGRA framework. In Ref.\cite{hesselb} it is shown
how precise determination of the light chargino and both selectron masses
could be used to determine the intra-generational slepton mass splitting
at the grand unification scale. This, in turn, will allow distinction
between the mSUGRA and the minimal gaugino mediation \cite{mgm}
framework for which the sparticle mass spectra are qualitatively very
similar. Measurements of third generation sparticle masses are
particularly interesting since these may contain information about the
Yukawa sector which may be otherwise difficult to obtain. For
instance, determination of  $m(\nu_{\ttau})$ with a
precision of 2-3\% could provide striking confirmation of tau
neutrino Yukawa interactions \cite{rhn}.

It is clear from these considerations that if sparticle masses can be
determined at the part per mille or even the subpercent level,
measurements at a LC would provide extremely stringent tests of the underlying
framework. In view of its potential importance, we felt that the
precision claimed in Ref. \cite{tesla} warranted a careful
re-examination. In this paper, we examine in detail the prospects of
measuring second and third generation sneutrino masses via the energy
dependence of the cross section. Within all SUSY models with lepton
flavour conservation these are produced only via $s$-channel $Z$
exchange and the cross sections are rather small, just a few GeV beyond
the production threshold.  The electron sneutrino case is somewhat
different because it typically has a much larger cross section since it
can also be produced via chargino exchange in the $t$-channel. Moreover,
selectron and $\tnu_e$ pair production (due to the much larger
production cross section) can be a significant source of SUSY
contamination for the $\tnu_{\mu}$ or $\tnu_{\tau}$ signal.  We find
that at least for the second and third generation sneutrinos, the event
rate in relatively background free channels is too small to allow a
threshold scan. Instead we propose an alternative strategy by which the
mass may be extracted and make projections for the precision with which
this might be possible.  While our focus is on electron and muon type
sneutrinos, for completeness we also examine the precision with which
$m(\tnu_e)$ might be obtained at a LC.

We base our results on the analysis of two mSUGRA case studies with
somewhat different kinematics and cascade decay patterns of sneutrinos.  The
first case is the one examined in the TESLA Technical Design Report
\cite{tesla} for which the parameters are,
\begin{displaymath}
{\bf Case \ I:} \ m_0=100 \ {\rm GeV}, m_{1/2}=200 \ {\rm GeV}, \tan\beta=3,
A_0=0, \mu >0.
\end{displaymath}
The second case that we choose was studied in Ref.\cite{rhn} to gain
some idea of how well the tau sneutrino mass could be extracted. The
corresponding model parameters are,
\begin{displaymath}
{\bf Case \ II:} \ m_0=150 \ {\rm GeV}, m_{1/2}=170 \ {\rm GeV}, \tan\beta=5,
A_0=0, \mu >0.
\end{displaymath}
Several sparticle masses along with relevant branching fractions are
shown in Table \ref{tab:cases}. We see that for both these cases $m_h$
the mass of the lightest neutral scalar in the Higgs boson sector is
well below the current LEP bound\cite{lephiggs} of 113~GeV so that it is
quite likely that these cases are experimentally excluded. The reason
that we have chosen these cases is to facilitate comparisions with the
earlier studies where the precision with which the sneutrino masses
could be obtained was also examined.\footnote{The value of $m_h$ is
quite irrelevant to our analysis.} We should mention a peculiar feature
of Case~I. Here, $m(\tz_2)$ is just slightly larger than $m(\ttau_1)$ so
that the decay $\tz_2 \to \ttau_1\tau$ competes with other three-body
decays. The decay patterns of $\tz_2$ are thus unusually sensitive to
the mass spectrum, and indirectly therefore, also to our choice
$m_t=175$~GeV.

The remainder of this paper is organized as follows. In the next
section, we compare the technique based on the energy dependence of the
cross section (of which the threshold scan is a particular case) with
that based on kinematic reconstruction of masses as in earlier studies
\cite{jlc,mur,munroe,zdr,snownlc}. We point out some issues that
potentially degrade the mass precision that will be attained in
experiments at a LC. In Sec.~\ref{sec3} we propose how one might
optimize an energy scan in relatively clean channels where the signal is
rate-limited so that a threshold scan is not possible because the signal
is tiny in the vicinity of the threshold. In
Sec.~\ref{sec4} we apply this method and assess how accurately
$m(\tnu_{\mu})$ and $m(\tnu_{\tau})$ might be determined in experiments
at a LC. We also examine the precision with which the
electron sneutrino mass might be measured.  In Sec.~\ref{sec5}, we
examine other potential channels for the extraction of $m(\tnu_{\tau})$
and $m(\tnu_{\mu})$ but find that these suffer from significant
backgrounds; in contrast, for $m(\tnu_e)$ we find that some improvement may be
possible, at least in favourable cases, by combining the signal on
several channels.  We present a summary of our
results along with general conclusions in Sec.~\ref{sec6}.

\section{Problems of determining sneutrino masses via an energy scan}
\label{sec2}

The extraction of the sneutrino mass from the energy dependence of the
cross section is essentially a counting experiment. We have either to
work in a channel that is relatively free of SM background and
contamination from other SUSY sources, or develop a procedure for
reliably subtracting these backgrounds. Generally speaking, the latter
would be preferable in that it allows for a larger signal (since we do
not have to limit ourselves to any particular channel) but we will see
later that both SM backgrounds as well as SUSY contamination can be
large. Moreover, in many SUSY models, all three generations of
sneutrinos are expected to be approximately degenerate; as a result, the
background from $\tnu_e\tnu_e$ production to second or third generation
sneutrino production will have essentially the same energy dependence,
and so will be difficult to subtract in a reliable manner using the data
below the signal threshold. In the following we will, therefore, mainly
focus our attention on the $\tnu_{\tau}\tnu_{\tau} \to
\tau\tw_1\tau\tw_1 \to \tau\tau jj\ell+\eslt$ ($\mu\mu jj\ell+\eslt$)
channel with $\ell=e,\mu$ where both SM backgrounds as well as SUSY
contamination to $\tnu_{\tau}$ ($\tnu_{\mu}$) pair production are
relatively small for a right polarized electron beam.\footnote{This is
the analogue of the $ee jj \mu$ channel used for the extraction of the
electron sneutrino mass from the energy distributions of electrons from
$\tnu_e \to e\tw_1$ decay \cite{munroe}. In this study, where it was
important not to confuse a lepton from chargino decay with the electron
from the primary decay of the sneutrino, this lepton was required to be
$\mu$. Here, since we do not need to identify the lepton from sneutrino
decay, we allow this to be either $e$ or $\mu$ to increase the signal.}
The inclusion of other channels results in substantial SM background
and/or SUSY contamination as discussed in Sec.~\ref{sec5}.

In Fig.~\ref{fig:bf}, we show contours of sneutrino mass as well as
those for the branching fraction for the decay chain,
$\tnu_{\tau}\tnu_{\tau} \to \tau\tw_1\tau\tw_1 \to \tau\tau
q\bar{q}\tz_1 \ell\nu\tz_1$ in the $m_0-m_{1/2}$ plane for $A_0=0$ and
$\mu >0$. This is the sign of $\mu$ favoured by the E821 experiment
\cite{bnl}. For this sign of $\mu$ the chargino tends to be lighter so
that there is more phase space for the cascade decay of the
sneutrino. We illustrate the contours for {\it a})~a low value of
$\tan\beta=3$ and {\it b})~ a high value of $\tan\beta=40$. Also shown
is the contour of $m_{\tw_1}=100$~GeV which is
roughly its lower mass limit~\footnote{$m_h$ is below the current LEP
bound especially for the $\tan\beta=3$ case, as we have already
mentioned.} from LEP experiments \cite{lephiggs}. The selectrons are
heavier than 100~GeV throughout both planes. 
The dark shaded region
in frame {\it b}) is excluded because $m^2(\ttau_R) < 0$, while the
light shaded region is disfavoured because $m(\ttau_1) < m(\tz_1)$.
The branching ratio
falls off to below 1\% for large values of $m_{1/2}$ because the decay
$\tw_1 \to \ttau_1\nu_{\tau}$ becomes dominant. 
We see that Case~II which has a branching fraction of
9.7\% for this cascade decay chain is very typical as long
as the two body decay of the chargino into the lighter stau is kinematically
forbidden. Case~I, while not atypical, has the corresponding branching
fraction towards the lower end of its range within this framework.

Longitudinal electron beam polarization is very effective in removing
both SM background as well as SUSY contamination
\cite{jlc,mur}. Polarization of the positron beam would also help (if it
is achieved without significant loss of luminosity) because the signal
cross section increases, but since the availability of polarized
positron beams appears less certain, we perform the bulk of our analysis
for unpolarized positron beams.  To identify potential sources of SUSY
contamination and also to show how these might be reduced, we show in
Fig.~\ref{fig:csections} the production cross sections for the most
important SUSY processes as a function of the electron beam polarization
parameter $P_L = f_L-f_R$, where $f_L$ ($f_R$) is the fraction of left
handed (right handed) electrons in the beam. These have been obtained
using ISAJET v7.51 \cite{isajet}. The two frames respectively show the cross
sections for ({\it a}) Case~I, and ({\it b})~Case~II introduced in the
previous section. Except for $\tmu_L\tmu_R$ production which has a
negligible cross section, the cross section for smuon pair production is
close to that for the corresponding third generation sparticle
production with $\ttau_1$ ($\ttau_2$) replacing $\tmu_R$ ($\tmu_L$). The
obvious point to note is that for signals of second or third generation
sneutrinos, a (dominantly) right handed electron beam reduces SM
backgrounds from $WW$ and $WWZ$ \cite{wwz} production as well as SUSY
contamination from the largest visible SUSY processes.\footnote{Since
$\te_R \to e\tz_1$, $\te_R\te_R$ production does not contaminate the
signal.} In the rest of our analysis, we fix $P_L(e^-)= -0.9$.

To get some idea of the signal rates near threshold, we show the number
of $eejj\ell$ events expected (before any cuts) from $\tnu_e\tnu_e\to
e\tw_1 e\tw_1 \to ejj+e\ell\nu +\eslt$ ($\ell=e,\mu$) production (solid)
and from $\tnu_{\mu}\tnu_{\mu}\to \mu \tw_1 \tz_2\nu \to \mu
jj+ee+\eslt$ production (dashed) for $P_L(e^-)=0.9$ in
Fig.~\ref{fig:thresh}{\it a}. For brevity, we show this only for Case I
but the results for Case II are qualitatively similar. The higher
sneutrino production cross section in Case I is partly compensated by
the fact that the branching ratio for the particular cascade decay chain
is correspondingly smaller. Following Ref. \cite{tesla}, we assume an
integrated luminosity of 10~fb$^{-1}$ for each value of $\sqrt{s}$, and
take the scan to extend to about 10~GeV above the threshold. As
discussed previously, the SUSY contamination exemplified by the dashed
curve is tiny (the threshold for $\tell_L\tell_L$ production is beyond
the range in the figure), and the signal rates high enough to imagine
carrying out a measurement of $m(\tnu_e)$ via a threshold scan as
suggested in Ref.\cite{tesla}.\footnote{$\tnu_{\tau}\tnu_{\tau}$
production leads to events with a $\tau$ rather than $\mu$ and so serves
to contaminate the signal only if the $\tau$ decays leptonically, so
that this background is below the dashed curve in
Fig.~\ref{fig:thresh}{\it a}. Moreover, with efficient vertex detection
this may be reduced even further.}  A potential difficulty with this
strategy is discussed shortly.

In Fig.~\ref{fig:thresh}{\it b} we show the corresponding expectations
for the $\tau\tau jj\ell$ channel, the favoured final state from tau
sneutrino pair production. The $\tnu_{\tau}$ signal cross section is
shown as the dotted curve, while the solid and dashed curves denote
contamination from electron and muon sneutrinos, respectively. Notice
that we have flipped the electron beam polarization. We see from
Fig.~\ref{fig:csections} that while there is slight reduction of the
$\tnu_{\tau}$ signal due to the polarization, the contamination from
$\tnu_e\tnu_e$ production is greatly reduced. For $P_L(e^-)=0.9$, we
have checked that the contamination from $\tnu_e$ pairs is an order of
magnitude larger than the stau signal!\footnote{This could be eliminated
by requiring $\ell=\mu$ but at a cost of a factor 2 in the already tiny
signal.} The most striking feature of the figure is that even before
experimental cuts and efficiencies are included, the signal yields just
a fraction of an event throughout the range of the scan. The
availability of positron polarization increases the cross section; for
$P_L(e^+)=0.6$ (corresponding to a positron beam with a dominantly
singlet component), the rate is higher by about 30-40\%, but clearly
this is still insufficient for our purpose.  Furthermore, even in this
``clean'' channel contamination from other sneutrinos is very
significant, and this contamination will be present in all SUSY models
where the different flavours of sneutrinos are approximately
degenerate.\footnote{We should remark that in the case of $\tnu_e$ or
$\tnu_{\mu}$ pair production, the taus dominantly come from decays of
$\tz_2$ produced via $\tnu \to \nu\tz_2$, so that $m_{\tau\tau}\leq
m_{\tz_2}-m_{\tz_1}$. It may, therefore, be possible to find cuts to
select out $\tnu_{\tau}$ events over $\tnu_e$ or $\tnu_{\mu}$ events at
some cost to the signal. However, since the signal is so tiny, it did
not make sense to explore this any further.}  Finally, we remark that
although we have not shown this explicitly, very similar considerations
will apply to the $\mu\mu jj\ell$ signal from $\tnu_{\mu}\tnu_{\mu}$
production.  We conclude that given the current projections for the
luminosity of a LC, the threshold scan does not seem to be a viable way
for precision measurement of second and third generation sneutrino
masses, primarily because the cross section in the relatively clean
channel is too small.

Although our discussion up to now makes it appear that it
should be possible to determine $m(\tnu_e)$ rather precisely via a
threshold scan, there is one potential difficulty that could significantly
degrade the precision from naive expectations~\cite{tesla}.
This arises because the cross section for any particular
final state depends on the sneutrino mass as well as the {\it a priori}
unknown branching fraction for the cascade of decays into the channel
being examined. This is not an issue for the lightest visible SUSY
particle which always decays directly to the lightest supersymmetric
particle (LSP), but is important for heavier sparticles which are the focus of
our analysis. We have checked that if we have measurements only near the
threshold, the freedom of the overall normalization greatly degrades the
precision, because the change in the cross section due to a slightly
different mass can be compensated for by a small change in the branching
fraction.  For instance, for Case~I with $m_{\tnu}=158.3$~GeV, even at
$\sqrt{s}=390$~GeV (which is quite far from threshold), the cross
section changes by $\sim 5$\% if the $m_{\tnu}$ is taken to be just
1~GeV larger than its reference value. Thus an observed event level
could be equally well fitted by the nominal mass and branching fraction
into the channel in question, or by a mass that is 1~GeV heavier and a
branching ratio that is 4.6\% rather than 4.4\%. Thus, {\it without
precise knowledge of the branching ratio, a determination of sneutrino
masses at the part per mille or even the subpercent level via a
threshold scan seems impossible, at least if backgrounds restrict us to
particular final states.} This state of affairs can be ameliorated by a
measurement of the event rate sufficiently far from threshold.  Here, we
assume the LC will first operate at $\sqrt{s}=500$ GeV, and that the
sneutrino threshold does not accidently lie too close to this.  A
measurement of the event rate in the continuum is much less sensitive to
$m(\tnu)$ but strongly constrains the branching fraction. The optimal
strategy, therefore, involves measurements in the continuum together
with measurements closer to the threshold.

A related (though possibly less severe) problem with the claims of very
precise determination of masses using the counting experiment strategy
is that the production cross section depends on other sparticle masses
({\it e.g.} of charginos in the case of $\tnu_e\tnu_e$ production or
neutralinos in the case of selectron production). Any uncertainty in
these will be reflected in the corresponding uncertainty in the
extraction of the selectron/sneutrino mass. For second and third
generation sneutrinos this is academic as we will see that their masses
cannot be extracted with subpercent precision, and further, that the
uncertainty due to the unknown branching fraction is far bigger. For the
extraction of the mass of $\te_R$ though (where there is no branching
fraction uncertainty) \cite{feng}, and potentially also for the much
better determined $m(\tnu_e)$, this could be a factor.

Before closing, we should mention that the uncertainty from the unknown
branching fraction has not been factored into earlier
analyses\cite{munroe,zdr,snownlc} of masses of sparticles that cascade
decay to the LSP, so that the precision may also have been
over-estimated in these studies. We may expect that this is a less
important factor in these studies since the end points of the energy
distributions are determined by  just the kinematics of the decay, and
are independent of the precise branching fraction for the decay chain.

\section{Optimizing the strategy for counting experiments with small signal
rates} \label{sec3}

The considerations of the previous section make it clear that for rate
limited final states such as $\tnu_{\mu}\tnu_{\mu} \to \mu\mu
jj\ell+\eslt$ or $\tnu_{\tau}\tnu_{\tau} \to\tau\tau jj\ell+\eslt$ the
strategy suggested in Ref. \cite{tesla}, {\it viz.} measuring the cross
section at ten points spaced 1~GeV apart starting just above the
threshold, is not feasible. First, the smallness of the cross section
necessitates that the available integrated luminosity be distributed
over fewer points, and second, we must have a measurement of the cross
section far above threshold. The goal of this section is to study how
a given integrated luminosity (which we take to be 120~fb$^{-1}$) ought
to be distributed to optimize the extraction of the sneutrino
mass. Since our purpose here is only to develop a suitable strategy,
for now, we assume the ideal case of 100\% detection efficiency and
geometric coverage for the detector and ignore backgrounds
from SM or other SUSY sources. Our considerations will thus apply
equally well to $\tnu_{\tau}$ and $\tnu_{\mu}$ since the production
rates and decay patterns are virtually identical for the small value of
$\tan\beta$ used in the analysis. In Sec.~\ref{sec4}, where we apply this
strategy, the effects of cuts, detection efficiencies and backgrounds
will be included.

The energy and sneutrino mass dependence of the cross section for $e^+
e^- \to \tnu_\tau \tnu_\tau \to \tau\tau jj\ell+\eslt$ can be written as,
\begin{equation}
\sigma(e^+ e^- \to \tnu_\tau \tnu_\tau) =
A \frac{s}{(s-M_Z^2)^2+M_Z^2 \Gamma_Z^2}
\biggl(1-\frac{4m_{\tnu_\tau}^2}{s}\biggr)^{3/2}\;,
\label{theory}
\end{equation}
where the normalization $A$ includes the products of branching fraction
for decays to the $\tau\tau jj\ell$ final state. We assume that the
available integrated luminosity ${\cal L}$ is divided up as ${\cal
L}={\cal L}_0+{\cal L}_{low} = 120$~fb$^{-1}$, where ${\cal L}_0$ is the
integrated luminosity at the nominal machine energy $\sqrt{s_0}$ which
we take to be 500~GeV, and ${\cal L}_{low}$ the integrated luminosity
divided up between $N$ lower energy points between the threshold and
$\sqrt{s_0}$. We assume that these points are equally spaced starting at
an energy $D$ above the production threshold, and parametrize their
locations by,
\begin{equation}
\sqrt{s_i} = 2 m_{\tnu_\tau} + D + (i-1)\Delta\;, \;\;i=1, ..., N.
\end{equation}
Our problem then is to choose ${\cal L}_0, D, \Delta$ and $N$ so as to
optimize the mass determination.

Choosing $D$ as small as possible seems to be intuitively optimal as
long as the event rate is sufficiently large. In our analysis, we
require that there be at least 10 events at each
point.~\footnote{Requiring just 10 events before efficiency and
acceptance corrections is somewhat over-optimistic, since we will see
later that once experimental acceptances and cuts are folded in, the
detection efficiency is 10-20\% (25-30\%) for $\tau\tau jj\ell$ ($\mu\mu
jj\ell$) events.}  This severely restricts $D$, because going too close to the
threshold results in a smaller number of events. We then distribute the
luminosity ${\cal L}_{low}$ over the $N$ points so that we expect about the
{\it same number of events at each of these points}. This ensures that
the cross sections at each energy are measured with similar precision,
but requires that more of the luminosity is spent close to threshold
relative to higher energies.

We then proceed as follows. For a particular choice of ${\cal L}_0, D,
\Delta$ and $N$, we generate a set of Monte Carlo data for model
parameters corresponding to Case~I so that for each value of
$\sqrt{s_i}$ the number of events in the ``data'' is a Gaussian
fluctuation about its theoretical expectation. Next, we fit this
``data'' to the theory (\ref{theory}), varying the sneutrino mass and
the branching ratio $BR$ (implicitly contained in the parameter $A$) into
the $\tau\tau jj\ell$ final state, and obtain the best fits to these two
quantities by minimizing $\chi^2$. The best fit values of $m_{\tnu}$ and
$BR$ do not, of course, coincide with their ``input values''. Finally, we
compute $\Delta\chi^2=\chi^2-\chi_{min}^2$ for different theories with
slightly different values of $m_{\tnu}$ and $BR$, and obtain the contour
with $\Delta\chi^2$=4.61 which yields the 90\% CL limits on how well
these parameters might be determined experimentally for the chosen
values of ${\cal L}_0, \Delta$ and $N$. The procedure is then repeated
to optimize this choice.

We found that choosing $N$ to be too large was far from optimal because the
event rate is small. The choice $N=10$ as in Ref.~\cite{tesla}
resulted in a
very poor determination of the mass parameter. The best results were
obtained for $N\leq 3$.
An illustrative example of our results is shown in
Fig.~\ref{fig:ellipse} for $N=3$, $\Delta=5$~GeV and {\it a})~${\cal
L}_0$= 60~fb$^{-1}$ (correspondingly, $D=45$~GeV), {\it b})~${\cal
L}_0$=40~fb$^{-1}$ ($D=32$~GeV), {\it c})~ ${\cal L}_0$= 20~fb$^{-1}$
($D=26$~GeV) and {\it d})~${\cal L}_0=0$ ($D=21$~GeV). The total
integrated luminosity has been taken to be 120~fb$^{-1}$. We see that for
the particular choice of $N$ and $\Delta$, frames
{\it b}) and {\it c}) show the best determination of the sneutrino mass. In
frame {\it a}) where ${\cal L}_0$= 60~fb$^{-1}$ the 10 event requirement
does not allow us to sample closer than 45~GeV from the threshold,
resulting in a degraded mass measurement. The results in frame {\it d})
confirm our earlier discussion: without a measurement in the continuum to
constrain the branching fraction, the mass cannot be well determined.

In Fig.~\ref{fig:ellipse}, we had arbitrarily fixed $\Delta=5$~GeV, a
rather small value to simulate an energy scan ``close'' to threshold. The
results of an analysis with variable $\Delta$ is shown in
Fig.~\ref{fig:delta}. Here, we have fixed $N=3$, ${\cal L}_0$ =
20~fb$^{-1}$ and taken {\it a})~$\Delta=20$~GeV with $D=19$~GeV, {\it
b})~$\Delta=40$ with $D=16$~GeV, {\it c})~$\Delta=60$~GeV with
$D=16$~GeV, and {\it d})~$\Delta=80$ with $D=15$~GeV. We see that the
precision on the mass is considerably improved relative to
Fig.~\ref{fig:ellipse}. This is mostly because for the larger values of
$\Delta$ in Fig.~\ref{fig:delta}, we can go closer to the threshold
because the cross sections at all but the lowest energy point are considerably
bigger, so that less luminosity is needed to achieve the 10 event level
at these intermediate points. Morover, we see that frames {\it c}) and
{\it d}) show a comparable precision on the sneutrino mass so that the
precise value of $\Delta$ is unimportant as long as it is sufficiently
large. An important conclusion is that {\it in a rate limited channel,
even allowing a measurement in the continuum, an energy scan close to the
threshold does not seem to be the optimal strategy for determining the
sneutrino mass.}

We have examined several choices of ${\cal L}_0, D, \Delta$ and $N$ to
optimize the strategy for extracting the sneutrino mass. We
summarize our findings here. For a good mass measurement:
\begin{enumerate}
\item It is necessary to go as close to the threshold as possible
consistent with the minimum event level. This requires more integrated
luminosity at the lower energy point where the cross section is the smallest.

\item It is necessary to have a reasonable cross section measurement
well above threshold to constrain the branching fraction. This comes
``for free'' because presumably the collider will first run at the
nominal energy which we have taken to be 500~GeV. We found that out of a
total integrated luminosity of 120~fb$^{-1}$, having about 20~fb$^{-1}$ at
$\sqrt{s}=500$~GeV was optimal.

\item In addition to the lowest energy point and the continuum point,
just one or two additional points are needed to get the mass. Having too many
points requires the total luminosity to be divided too much resulting in
a loss of precision.

\item The precise interval between the points scanned is not very
important. It is important, however, that the interval be large enough
so that these points are not all concentrated very close to threshold
where the cross section is small, and one does not obtain any
measurement sufficiently close to the threshold.
\end{enumerate}.

In our study, the results in Fig.~\ref{fig:delta}{\it c} and {\it d}
yield about the best measurement of the sneutrino mass. We thus conclude
that with an ideal detector and no background, as well as no initial
state radiation, beam energy spread or beam-beam interaction effects, the
tau sneutrino mass can be determined at about the $\sim 3$\% level with
an integrated luminosity of 120~fb$^{-1}$, unless
the branching fraction for the decay cascade can independently be
determined from other measurements.\footnote{One might think that
reducing $N$ to 1 would allow a better measurement since $D$ could be
chosen to be smaller still. While this may be true, this is not really
possible in practice since one would have to know the position of the
threshold quite precisely to reduce $D$ significantly.}  Even under
these idealized circumstances, our conclusions about the precision on
the mass that can be attained are considerably more pessimistic than
those in Ref. \cite{tesla}. We see, however, that the energy scan also
fixes the branching fraction for the decay to the particular final state
to be between 3-6\%. It is worth remarking that the branching fraction
is slightly better constrained if ${\cal L}_0$ is chosen to be larger,
as for example, in Fig.~\ref{fig:ellipse}{\it a}.  Before closing this
discussion, we should mention that we have ignored effects of the
sneutrino widths in our analysis. Since $\Gamma_{\tnu}$ is typically a
fraction of a GeV, while the smallest value of $D$ is larger than
15~GeV, non-zero width effects \cite{widths} are completely negligible.

\section{Realistic determination of sneutrino masses} \label{sec4}

In this section we use the energy scan strategy devised in
Sec.~\ref{sec3} to investigate the precision with which the tau and muon
sneutrino masses might be measured in LC experiments once effects of
finite detector acceptances, experimental cuts, SM backgrounds, SUSY
contamination, and finally, beam energy smearing
due to initial state radiation (ISR) and beamstrahlung are incorporated.

\subsection{Beamstrahlung and Initial State Radiation}

The main effect of beamstrahlung and ISR is to reduce the energy in the
electron/positron beams due to emission of photons, thereby reducing the
available centre of mass energy in the $e^+e^-$ collision.  Since our
strategy entails a measurement of the event rate as close to threshold
as possible, and the cross section varies rapidly with energy near the
threshold, these effects are especially important. We include these
effects following the suggestions in Ref. \cite{peskin}. For ISR, we use
the structure function approximation suggested by Skrzypek and Jadach
\cite{sj}. We include beamstrahlung by using the electron energy
spectrum as given by Peskin \cite{peskin}. We set the parameter
$N=N_{\gamma}/2$ with $N_{\gamma}= 1.176$ and take $\Upsilon = 0.124$. This
choice corresponds to the design parameters in Ref. \cite{nlc} for
$\sqrt{s}=500$~GeV.

The reduction of $\sigma(\tnu_{\tau}\tnu_{\tau})$ (or
$\sigma(\tnu_{\mu}\tnu_{\mu})$) is illustrated in Fig.~\ref{fig:beam}. The
reduction, which is shown relative to the nominal cross section
$\sigma_0$, is a function of just $\sqrt{s}/2m$, where $m$ is the
sneutrino mass. The diamonds and triangles show the result for Case~I
and Case~II, respectively, while the line is a fit through the points.
We see the striking reduction of the cross section close to the
threshold as expected. We will fold this reduction into our evaluation
of the energy dependence of the cross
section, and into our assessment of the precision with which the
sneutrino mass might be measured. We should mention that this
is perhaps too conservative since we will use the beamstrahlung
parametrization appropriate to a 500~GeV collider all the way down to about
350~GeV where the beamstrahlung and ISR effects, and hence the reduction
of the cross section, are smaller.

\subsection{Monte Carlo Simulation}

We use ISAJET v7.51 for our SUSY event simulation as well as simulation
of $2 \to 2$ SM backgrounds.
We use a toy calorimeter covering
$-4<\eta <4$ with cell size $\Delta\eta\times\Delta\phi =0.05\times 0.05$.
Energy resolution for electrons, hadrons and muons is taken to be
$\Delta E=\sqrt{.0225E+(.01E)^2}$, $\Delta E=\sqrt{.16E+(.03E)^2}$ and
$\Delta p_T =5\times 10^{-4} p_T^2$, respectively.
Jets are found using fixed cones of size
$R=\sqrt{\Delta\eta^2+\Delta\phi^2} =0.6$ using the ISAJET routine
GETJET (modified for clustering on energy rather than transverse energy).
Clusters with
$E>10$ GeV and $|\eta ({\rm jet})|<2.5$ are labeled as jets.
Muons and electrons are classified as isolated if they have $E_T>5$ GeV,
$|\eta (\ell )|<2.5$, and the visible activity within a cone of $R=0.5$
about the lepton direction is less than
$max({E_{T\ell}\over 10},\ 1\ {\rm GeV})$. Identification of taus is
discussed below.
In addition to these basic acceptance cuts, we also require $\eslt \geq
25$~GeV to enhance the SUSY signal.

\subsection{Tau Sneutrinos}

We focus here on the $\tau\tau jj\ell$ signal that we have been
discussing. Here $\ell$ is defined to be an electron or muon not tagged
as coming
from tau decay. The cross section in this channel is sensitive to how
well taus can be identified.  Usually it is assumed that only
hadronically decaying isolated taus can be identified as narrow jets
with one or three charged tracks\footnote{A track is required to have a
$p_T$ of at least 0.5~GeV.} with total charge $\pm 1$ and $m_{tracks}
\leq m_{\tau}$, with little hadronic activity around them. This is
implemented via a cone algorithm \cite{rhn} that required the tracks to
be in a 10$^\circ$ cone about the jet axis, with no additional hadronic
activity in the corresponding 30$^\circ$ cone.  For identification of
both taus in the $\tau\tau jj\ell$ channel, we then immediately lose a
factor 4/9 relative to the corresponding signal for muon
sneutrinos. Since the signal in this channel is already small, it is
worthwhile to seriously consider any possibility for more efficient
$\tau$ identification.
Following our discussion in Sec.~\ref{sec1}, we will optimistically
assume that {\it all} leptonically decaying taus (where the secondary
leptons satisfy $p_T$ and geometric cuts of the previous subsection) are
tagged. ``Tau jets'' would also have a displaced vertex, but this is
also true of heavy flavour jets.  Moreover charm has a lifetime and mass
similar to tau, so that charm quarks from the decays $\tw_i \to cs\tz_1$
and $\tz_j \to c\bar{c}\tz_1$ would contaminate the tau sample. A
$c$-quark would fragment into a $D^{(*)}$ meson which would rapidly
decay to a (weakly decaying) $D$ meson and additional pions or kaons;
this lightest of the $D$ mesons would decay weakly, resulting in a
displaced vertex due to its long lifetime. If the fragmentation products
and the pions/kaons from the strongly decaying $D^{(*)}$ meson are soft,
or away from the final weakly decaying $D$ (which has a mass close to
$m_{\tau}$), the $c$ jet can mimic a hadronically decaying $\tau$
lepton.  Even if it is possible to efficiently discriminate between
$\tau$s and $b$s via the difference in their masses and lifetimes, we
need to examine whether additional discrimination between tau and charm
jets is needed.

Toward this end we have computed the cross section for $4j+\ell$ final
state from all SUSY sources, where one of the four jets comes from a
hadronically decaying $\tau$ and at least one of the remaining three is
a charm jet. We find that
this cross section is of the same size as the $\tau\tau jj\ell$ signal
cross section from hadronically decaying taus, so that with vertexing
alone, $c$ jets would significantly contaminate the $\tau$ sample making
precision measurements impossible. Fortunately, unlike $\tau$ jets, most
$c$-jets are not expected to have just three tracks in the 10$^\circ$
cone with no tracks in the larger 30$^\circ$ cone.

To minimize unnecessary loss of the already small $\tau\tau jj\ell$
signal, rather than use the stringent cone algorithm of Ref. \cite{rhn}
for the identification of hadronically decaying $\tau$s and veto jets
with a track in the outer cone, we have examined the ratio,
\begin{equation}
r \equiv \frac{\sum_{i=1}^{N^{in}} E^{in}_i}{\sum_{i=1}^{N^{in}} E^{in}_i +
\sum_{j=1}^{N^{out}} E^{out}_j}\; ,
\label{eq:r}
\end{equation}
where $N^{in}$, which is required to be 1 or 3 and $N^{out}$ are the
number of tracks in the 10$^\circ$ and the 10$^\circ$-30$^\circ$ cones,
respectively. The cone algorithm used earlier requires $N^{out}=0$; {\it
i.e.} $r=1$. The dashed (solid) histogram in Fig.~\ref{fig:r}{\it a}
shows distribution of $r$ for real $\tau$-jets from
$\tnu_{\tau}\tnu_{\tau}$ events for Case~I (Case~II). The last bin
(beyond $r=1$) shows the event rate for $r=1$ and contains about 85\% of
the events. Since we require two taus in our signal, requiring $r=1$
reduces the cross section from hadronically decaying taus by about 30\%.
The corresponding distribution from QCD jets in the
$\tnu_{\tau}\tnu_{\tau}$ sample which coincidentally have $N^{in}= 1$ or
3 is shown in Fig.~\ref{fig:r}{\it b}. As expected, the distribution is
broad because typically QCD jets would have many more tracks than 3 and
some of these would lie in the outer cone. The peak at $r=1$ is
presumably either because the jet has only one or three tracks or, by
chance, the other tracks happen to be outside the 30$^\circ$ cone.

Assuming the distribution of tracks in $c$ jets is qualitatively similar to
that in Fig.~\ref{fig:r}{\it b}, it is clear that requiring $r \geq
r_{min}=0.5-0.8$ will reduce the charm contamination with little effect
on the signal. We have checked that even with $r \geq 0.5$, the
contamination from charm jets in SUSY events is reduced to about
20\% of the $\tau\tau jj\ell$ signal from just the hadronically decaying taus.
We stress, however, that we view the results of the
$\tau$-$c$ discrimination analysis as qualitative. A serious analysis of
this should include the $r$ distribution from just $c$-jets, effects of
differences in $\tau$ and $D$ meson lifetimes as well as incorporating the
(presumably small) contamination from $b$ and light quark jets. Morever,
the results may also be sensitive to the details of the quark
fragmentation as well as hadronization. The important message from our
discussion, however, is that vertex detection by itself {\it is not sufficient}
to eliminate contamination of $\tau$ signals, but with additional
requirements this can probably be effectively reduced. We will,
therefore, not consider it in the remainder of our analysis.

The importance of being able to tag $\tau$s via their leptonic decays is
illustrated in Table~\ref{tab:tau}. We show the results for three values
of the $c$-$\tau$ discrimination parameter $r_{min}$. Here $j^{\tau}$
($\ell^{\tau}$) refer to a tau identified via its hadronic (leptonic)
decay. The table shows the gain in the case of the hadronically decaying
$\tau$s from relaxing the original cone algorithm and including
$\tau$-jets with $r < 1$. The result differs somewhat from simple
expectations based on Fig.~\ref{fig:r} because while {\it all} taus were
included in the figure, here we only include events with two identified
taus, two jets and a lepton; {\it i.e.} the environment of the taus is
somewhat different. The main point of Table~\ref{tab:tau} is that if
vertex detection allows for efficient tagging of leptonically decaying
taus, the signal essentially doubles (and increases by a factor of 3
relative to that using the original cone algorithm). In the following we
will use the relaxed cone algorithm with $r_{min}=0.5$, and assume that
leptonically decaying taus can be efficiently tagged.

We are now ready to discuss the prospects for the determination of
$m(\tnu_{\tau})$.  Our procedure for this is essentially the same as in
Sec.~\ref{sec3} except that we now include the effects of ISR and
beamstrahlung, finite detector acceptances and cuts, and
backgrounds. To avoid repeated, lengthy simulations of the signal for
each point in the sneutrino mass vs branching fraction plane, we have
adopted the following procedure.
\begin{enumerate}
\item First, we obtain the total cross section into the $\tau\tau jj\ell$
final state. This step does not entail any event simulation.

\item Second, we correct this cross section for beamstrahlung and ISR
effects using the curve in Fig.~\ref{fig:beam}.

\item Third, we multiply the corrected cross section by an efficiency
function to take into account the effects of detector acceptance,
measurement resolution and experimental cuts. To obtain this we use
ISAJET to simulate $\tnu_{\tau}\tnu_{\tau}$ events for
SUSY parameters corresponding to the two cases that we have studied. We
assume that the efficiency changes slowly over the range of sneutrino
masses in our scan, and take it to be a constant.
\end{enumerate}
This procedure allows us to scan the signal in the $m_{\tnu}-BR$ plane
with just one signal simulation for each of the two cases. Of course,
the SUSY contamination has also to be simulated in each case as
discussed below.

The detection efficiency for the $\tau\tau jj\ell$ signal is shown in
Fig.~\ref{fig:taueff}. The diamonds (triangles) show the results of our
simulation for Case~I (Case~II) while the curves, which we will use for
our subsequent analysis are a fit. The efficiency varies between about
10-20\%. One might think that the main reason for this low efficiency
is due to inefficient tau detection, but this is not entirely the
case. Requiring five objects inside the detector acceptance together
with $\eslt \geq 25$~GeV already reduces the efficiency to 25-30\%
(see Fig.~\ref{fig:mueff} below). The additional reduction is indeed due
to differences in tau and muon detectability. The efficiency is
significantly smaller in Case~I, presumably because the
mass difference
$m(\tnu_{\tau})-m({\tw_1})$ is considerably smaller, resulting in
somewhat softer taus than in Case~II. The turnover for larger values of
$\sqrt{s}$ in Case~I is probably due to the fact that sneutrinos are
boosted so that their decay products tend to emerge closer in space, and
so find it harder to pass the isolation requirements.

In Fig.~\ref{fig:tauback} we show the $\tau\tau jj\ell$ signal along
with contamination from various SUSY sources for {\it a})~Case~I, and
{\it b})~Case~II after all the cuts as well as beamstrahlung and ISR
corrections. SM backgrounds from $2\to 2$ processes are negligible.
$e^+e^- \to WWZ$ production, where $Z \to \tau\tau$ and the $W$s each
decay to $jj$ and $\ell\nu$ is a potentially important background. For
an electron beam with $P_L(e^-)=-0.9$, the production cross section at
$\sqrt{s}=500$~GeV is just 4.8~fb (in contrast to 40~fb for an
unpolarized beam). Folding this with the branching fraction $8.9\times
10^{-3}$ for this decay chain yields a background level of just 0.04~fb
before any acceptance and identification cuts. Recalling the low
detection efficiency for the final state, we conclude that this
background is also negligible.
The
signal cross section, shown by the solid line, is indeed very small,
primarily because of the low detection efficiency. We have separately shown the
contamination from the first two generations of sneutrinos and sleptons
(circles), $\ttau_1\ttau_2$ and $\ttau_2\ttau_2$ production (triangles)
and chargino and neutralino production (squares). In the last case, at
least one of the charginos or neutralinos is $\tw_2$ or $\tz_{3,4}$. The
reason for separating out the contamination in this manner is that these
relative contributions will be model and parameter dependent.  In most
models (unless $\tan\beta$ is very large) we expect that $\te_L$,
$\tmu_L$, $\ttau_2$ and all three flavours of sneutrinos are
approximately degenerate (right handed sleptons and $\ttau_1$ pair
production does not contribute to the background) and have about the
same threshold as the signal. The masses of the heavier charginos and
neutralinos are not as directly correlated to $m(\tnu_{\tau})$ so that
the background shown by squares will depend on the model as well as on
the choice of parameters. Indeed we see this in the figure. In Case~I,
this background becomes important only at the highest energy, while in
Case~II, this is the main background. It may even be that the production
of heavy charginos and neutralinos is kinematically inaccessible in a
particular model. The dashed line shows the total SUSY contamination.
We see that even in this ``best channel'' the signal to background ratio
is ${\cal O}(1)$.

The low signal cross section in Fig.~\ref{fig:tauback} means that it is
not possible to obtain a mass measurement with just 100~fb$^{-1}$ of
integrated luminosity.  Even at $\sqrt{s}=500$~GeV, just 2-5
signal events would be expected for an integrated luminosity of
20~fb$^{-1}$ for these cases. We find that for a reasonable mass
measurement an integrated luminosity of at least 500~fb$^{-1}$ is
necessary. For the remainder of this discussion, we will assume that a
data sample of this size will be available. This may be possible with as
little as two years \cite{tesla} to a few years of LC operation at the
design luminosity.

We divide this integrated luminosity into ${\cal L}_0$, which we take to be
100~fb$^{-1}$, and ${\cal L}_{low}$ which, as discussed in
Sec.~\ref{sec3}, we further divide amongst $N=3$ points to optimize the
measurement.  Our final result, the 90\% CL contour in the
$m(\tnu_{\tau})$ branching ratio plane is shown in
Fig.~\ref{fig:tauellipse} for {\it a})~Case~I, and {\it b})~Case~II.  We
have required that there be at least six events (after all cuts and
beamstrahlung/ISR corrections) at each point.  The solid line shows the
90\% CL contour assuming that there is no SUSY contamination, while the
outer dashed contour is obtained including the background in
Fig.~\ref{fig:tauback}. The cross shows the result for the best fit for
the solid contour; {\it i.e.} without the background included.  Several
comments are in order.
\begin{itemize}

\item The ``error ellipses'' from the counting experiment are asymmetric
in mass\footnote{Indeed the fact that the $\Delta\chi^2=4.6$ contours
are not ellipses should warn us to view the confidence interval from
this as a qualitative indicator since we are clearly in the non-Gaussian
regime.} in that the lower bound on the mass is further away from the
best fit value than the upper bound. In frame {\it a}) this asymmetry is
extreme. While it may be somewhat surprising at first, the asymmetric
error ellipse is readily understandable. The point is that if we try to
fit the ``data'' with too high a sneutrino mass, a bad fit is obtained
because some of the data points are quite close to the threshold for
this fitted mass, resulting in a large $\chi^2$ and a bad fit. On the
other hand, if we attempt to fit with too small a sneutrino mass, the
data are much more in the ``continuum region'' for this value of the
sneutrino mass, and there is no data point near the threshold for this
mass, so that the fit is not as poor, resulting in an asymmetric
ellipse. This argument suggests that no matter how much the integrated
luminosity is, there will be a (usually disjoint) region with a small
$m(\tnu)$ and a small branching ratio to compensate the cross section,
where the counting experiment alone gives a ``good fit'' to the
data. This would not be a problem if the sneutrino mass in this region
is smaller than $m_{\tw_1}$ which will be measured
\cite{jlc,mur,munroe,zdr,snownlc} to 1-2\% precision by other
techniques: sneutrinos with a mass below $m(\tw_1)$ clearly cannot give
the signal we are considering. Our examination of Case~I is an extreme
case of this phenomenon. The small signal cross section has caused the
lower region to merge with the original ``ellipse'', but it is not
meaningful to think that the sneutrino mass can be smaller than
$m(\tw_1)=129$~GeV.

\item We see that the inclusion of the background indeed reduces the
precision with which the mass and the branching ratio can be
determined. For the reason just discussed, the background increases the
error in the mass more toward the lower end than the upper.

\item We should keep in mind that we have examined just two cases. While
the SM background is negligible, the SUSY contamination will be model
dependent as noted above. Moreover, once the data
sample is in hand, a detailed examination of the events may make it
possible to reduce the contamination from charginos and neutralinos and
possibly also from the first two generations of sneutrinos and sleptons
since in these cases the taus are produced from chargino and neutralino
decays, while in the signal they result from the primary decay of the
produced sparticle. It seems fair, therefore, to say that the precision
that might be attained is somewhere between the dashed and solid contours.

\item In obtaining the error ellipses, we have taken the SUSY
contamination to depend only on the energy; {\it i.e.} we compute this
for the input parameters of the model, and assume that it does not
change as we scan the $m_{\tnu}-BR$ plane. This is, of course, not true
in a particular framework such as mSUGRA or gauge-mediated SUSY breaking
where all sparticle masses and couplings are determined by just a few
parameters. In these cases, the data would be used to determine these
underlying parameters which would then be used to fix for instance
$m(\tnu_{\tau})$. Indeed in particular frameworks, it is entirely
possible that the tau sneutrino mass may be much better measured than
shown in Fig.~\ref{fig:tauellipse}, since data from $\tnu_e\tnu_e$
production, which has a two order of magnitude larger cross section, can
be used to constrain even $m(\tnu_{\tau})$.  Another way of saying this
is that we treat the SUSY contamination as though we are in the
MSSM, and (for the purpose of analyzing measurements of
$m(\tnu_{\tau})$) we keep all other sparticle masses fixed. Again we
stress that while our analysis suggests that $m(\tnu_{\tau})$ may only
be determined to lie within a $\pm 8$~GeV ($\pm 11$~GeV) range without (with)
backgrounds in Case~II {\it a model-dependent analysis within a
particular framework may yield a much better result.}

\item We have checked that our results are quite insensitive to the
value of $N$, as long as it is small. Increasing $N$ to five did not
degrade the result much because the best results were obtained for a
value of $D$ which was essentially the same as for $N=3$. The additional
points were at relatively high energy where the integrated luminosity
for six events was not very large.

\item The dotted contour shows how the solid contour in
Fig.~\ref{fig:tauellipse} would shrink if we increase both ${\cal L}_0$
and ${\cal L}_{low}$ by 50\%. This roughly corresponds to an increase in the
cross section that would be obtained with a positron beam
polarization\footnote{The SUSY contamination also depends on the
positron beam polarization and would have to be computed in order to see
how the dashed contour is altered.}  of just over 60\%. In frame {\it
a}) we see the
emergence of the additional dotted region below $m(\tnu_{\tau})=90$~GeV,
in accord with our earlier discussion.

\end{itemize}
We conclude that the energy dependence of the cross section at best
allows a measurement of $m(\tnu_{\tau})$ with a precision of a few
percent at the 90\% CL. The branching fraction for the decay chain is
also constrained.

\subsection{Muon Sneutrinos}

The analysis for muon sneutrinos proceeds exactly along the same lines
as that for $\tnu_{\tau}$ just discussed. The efficiency \footnote{There
is an ambiguity in how we define this efficiency because the decay chain
$\tnu_{\mu}\tnu_{\mu} \to \mu\tw_1+\nu\tz_2 \to \mu jj+\mu\mu+\eslt$ also
leads to the $\mu\mu  jj\ell$ final state with $\ell=\mu$. Here, we have
defined the efficiency to be what would be obtained assuming the
branching ratio is that given just by the decay chain where both
sneutrinos decay via their chargino mode.} for detecting
$\mu\mu jj\ell$ events from the chain $\tnu_{\mu}\tnu_{\mu} \to \mu\tw_1
\mu\tw_1 \to \mu\mu jj\ell +\eslt$ is shown in Fig.~\ref{fig:mueff} for
the two cases we have been discussing. Again diamonds show the results
for Case~I while triangles show that for Case~II. We see that this
efficiency, which typically increases with energy, is about 25-35\% for
these scenarios.

Fig.~\ref{fig:muback}{\it a} and Fig.~\ref{fig:muback}{\it b} show the
signal from $\tnu_{\mu}\tnu_{\mu}$ production (solid) along with the SUSY
contamination from all charged sleptons and other sneutrinos (circles),
and from chargino and neutralino production (squares) for Case~I and
Case~II, respectively. We do not
separately show the background from smuons since this has essentially the
same threshold as selectrons in most models. The total SUSY
contamination is shown as the dashed curve. Except for the fact that the
cross sections are higher than in Fig.~\ref{fig:tauback} by a factor
2-3, the two figures are qualitatively very similar. The SM background
is again negligible.

The larger efficiency for muon sneutrino events should allow a better
determination of $m(\tnu_{\mu})$. This is borne out by the error
ellipses shown in Fig.~\ref{fig:muellipse} for {\it a})~Case~I, and {\it
b})~Case~II. The solid (dashed) curve shows the result without (with)
SUSY contamination. We see that in this case, the impact of the
background is much smaller than in our study of $\tnu_{\tau}$. This is
presumably because the larger cross section results in a higher
statistical significance of the signal. As in Fig.~\ref{fig:tauellipse},
the mass error is asymmetric, but the ellipse is much less skewed
primarily because of the higher event rate. Our study shows that
assuming that the branching fraction for the decay chain is not
constrained from elsewhere, $m(\tnu_{\mu})$ can be determined to lie
between about 148~GeV and 160~GeV (171~GeV and 180~GeV) for Case~I
(Case~II), suggesting that counting experiments may determine it with a
precision of about $\pm(2.5-4)$\% in a model-independent manner. The
branching fraction for the cascade sequence is determined within about
$\pm 50$\% though, once again, with an asymmetric error. This extracted
``branching ratio'' should, however, be interpreted with care since
other decay chains, {\it e.g.} $\tnu_{\mu}\tnu_{\mu} \to \mu
\tw_1+\nu\tz_2 \to \mu jj+\mu\mu+\eslt$ also leads to the $\mu\mu jj\ell$
final state (with $\ell=\mu$).

\subsection{Electron Sneutrinos}

Although an analysis of how well the electron sneutrino mass can be
determined is not the main purpose of this paper, it is a natural
extension of the present work. We should state at outset that our study
of this is mainly to give the reader
an idea of how much the issues raised above, which were very important
for the second and third generation sneutrinos, impact on the
determination of $m(\tnu_e)$. In this case, we do not mean to imply that
our study is optimized: indeed in the next section, we will suggest some
ways by which the precision may be improved.

For a study of the electron sneutrino, it is clear from
Fig.~\ref{fig:csections} that the electron beam should be polarized to
be mostly left-handed. In this case, it is clear that well above the
production threshold, the cross section is so large that SM backgrounds
as well as SUSY contamination are negligible. What is, however, not
clear is whether this will continue to be true close to the electron
sneutrino production threshold where the signal becomes small.  Since
neither SM backgrounds nor the production of lighter charginos or
neutralinos is kinematically suppressed at $\sqrt{s} \sim 2m(\tnu_e)$,
at least some of the channels are likely to be contaminated from these
sources. In keeping with the analyses for second and third generation
sneutrino masses we will, therefore, first focus on the $ee jj\ell$
channel which does not suffer from these backgrounds\cite{munroe}, and
defer the question of inclusion of other channels to the next section.

There are two questions concerning the strategy for determining
$m(\tnu_e)$ that we attempt to answer here. Except for the point at
$\sqrt{s}=500$~GeV that is essential to constrain the branching ratio,
is it better to scan several points near the threshold, or is it better
that these points be significantly spaced (large $\Delta$) so that we do
not spend most of the luminosity where the signal is small? Second, is
it better to divide the luminosity equally between the low energy
points, or is it best to apportion the luminosity so that we have about an
equal number of events for each energy point?

To answer these, we have analysed four distinct strategies for the
determination of $m(\tnu_e)$.  As before, we have first obtained
$\sigma(\tnu_e\tnu_e)$ corrected for beamstrahlung and ISR as before,
and then multiplied it by an efficiency function which differs somewhat
from that for smuon neutrinos because of the $t$-channel contribution
for $\tnu_e\tnu_e$ production.  We assume there
are no SM or SUSY backgrounds to the signal, and require at least six
signal events at any energy point.\footnote{For $\tnu_{\mu}$ and
$\tnu_{\tau}$ signals, this restricted how close we could go to the
threshold. In this case, this is not so for 500~fb$^{-1}$ of integrated
luminosity. However, to be able to be very close to the threshold, we
already need to have a good idea about $m(\tnu_e)$. In our analysis we
have conservatively assumed that $D \geq 2$~GeV, which should be quite
feasible since measurements in the continuum will already pin
$m(\tnu_e)$ with a precision of about 1\%
\cite{munroe,zdr,snownlc}. This also ensures that our neglect of
sneutrino widths does not cause too large an error.}

Our results for the 90\% CL ($\Delta\chi^2=4.6$) error ellipses in
the $m(\tnu_e)-BR$ plane are shown in Fig.~\ref{fig:eellipse} for Case~I
(first column) and Case~II (second column) for an integrated luminosity
of 120~fb$^{-1}$ (outer ellipses) and 500~fb$^{-1}$ (inner ellipses). We
divide the luminosity ${\cal L}_{low}$ between $N=3$ points and examine
the precision for both $\Delta=1$~GeV (solid ellipse), and
$\Delta=30$~GeV (dotted ellipse). As in the previous studies, we take
${\cal L}_{low}$ to be 100~fb$^{-1}$ (400~fb$^{-1}$). We perform this
analysis first by dividing ${\cal L}_{low}$ so that there are an equal
number of events for each of the three points (first row), and also
equally amongst the three low energy points. Thus, the solid ellipses in
the lower left frame correspond to the result of a conventional threshold
scan with three points spaced by 1~GeV, together with the point at
$\sqrt{s}=500$~GeV.
We note the following:
\begin{enumerate}
\item $\Delta=1$~GeV yields a more precise determination of $m(\tnu_e)$
than $\Delta=30$~GeV, especially in row 2, where the luminosity is
equally divided between the points. This is a reflection of the fact
that it is important to get a good determination of the cross section
close to the threshold. In the first row, the difference
between the dotted and solid ellipses is small, because most of the luminosity
is already spent at the threshold where the cross section is much
smaller.

\item As before (and for the same reasons), $m(\tnu_e)$ has an asymmetric
error with the 90\% CL range extending further to the lower side. For an
integrated luminosity of 120~fb$^{-1}$, $m(\tnu_e)$
can be determined to lie in a 2~GeV (1.3~GeV)
range for Case~I (Case~II). If an integrated luminosty of 500~fb$^{-1}$
is available, the range shrinks by a factor of about 2.

\item With 120~fb$^{-1}$ of integrated luminosity, the branching
fraction for the decay cascade can be determined to better than $\pm
10$\% for both cases, again with an asymmetric error. The determination
of branching fraction and mass are somewhat complementary since the
strategy that yields the most precise value of $BR$ yields the largest
uncertainty in mass. A data sample of 500~fb$^{-1}$ will reduce the
error on the branching fraction by about a factor 2. As in the muon
case, the extracted branching ratio must be interpreted with care, since
this final state can also arise via other decay chains.
\end{enumerate}
We thus conclude that unless other channels can be included, the
sneutrino mass may be determined with a precision of $\sim 0.5$\%, the
exact value depending
on the branching fraction with an integrated luminosity of
120~fb$^{-1}$. The precision will about double if an integrated luminosity of
500~fb$^{-1}$ is available.

Since we are talking about a sneutrino mass measurement at the sub-GeV
level from counting rate alone, it is necessary to ask whether
theoretical uncertainties in the sneutrino production cross section
could vitiate such a claim. If this production cross section is uncertain
by a considerable amount, this would effectively reflect itself as an increased
error in
the branching fraction, which would result in a degradation of the mass
measurement. As long as the relative error in the cross section remains
smaller than that in the branching fraction from the error ellipses of
Fig.~\ref{fig:eellipse}, we expect that this will be a subdominant
effect, and our previous conclusions will remain valid.

Uncertainties in the cross section could arise due to limitations in our
knowledge of masses (or mixing angles) of charginos on which the
production cross section depends \cite{feng}. We should keep in mind
that by the time experiments at LCs are ready to do sub-GeV measurements
of $m(\tnu_e)$ the mass of $\tw_1$ will be determined to better than
about 1\% \cite{jlc,mur,munroe,zdr,snownlc}.  We will assume that the
heavier chargino mass, assuming (as for the case studies in this paper)
$e^+e^- \to \tw_1\tw_2$ production is kinematically accessible, is known
with a precision that is no more than five times worse. To understand
the impact of these uncertainties in $m(\tw_1)$ and $m(\tw_2)$ on the
electron sneutrino production cross section, we have computed the change
in this cross section by altering $m(\tw_1)$ and $m(\tw_2)$ from their
nominal values, with the signs of $\Delta m(\tw_i)$ chosen to maximize
this change.  We show the fractional uncertainty in the cross section as a
function of $\Delta m(\tw_1)/m(\tw_1)$ the relative precision with which
it is measured, assuming that $\Delta m(\tw_1)/m(\tw_1)= 5\Delta
m(\tw_2)/m(\tw_2)$. The result of our computation of the {\it maximum}
possible change in the cross section (due to mismeasurements of chargino
masses) is shown in Fig.~\ref{fig:unc} for
Case~I (Case~II) by diamonds (triangles) for {\it a})~$\sqrt{s}=500$~GeV,
and {\it b})~$\sqrt{s}=2m(\tnu_e)+2$~GeV. The dot-dashed (dotted) band
at 2.7\% (4.5\%) corresponds to the uncertainty of the cross section due
to the uncertainty in the branching ratio (with 500~fb$^{-1}$ of
integrated luminosity) from the error ellipses in
Fig.~\ref{fig:eellipse}. We see that as long as the light chargino mass
is measured with a precision better than 1.5\% (2.5\%) in Case~I
(Case~II), the uncertainty from an imperfect knowledge of the chargino
masses is smaller than that resulting from the error in the branching
fraction in Fig.~\ref{fig:eellipse}, and our previous conclusions about
the precision that would be possible are not greatly altered.

If the heavier chargino is kinematically inaccessible so that $m(\tw_2)$
cannot be measured, this uncertainty
might be larger. For any model with just two charginos, the extreme
limit would be $|\mu| \to \infty$ with $M_2$ being adjusted to keep
$m(\tw_1)$ within its measured range. In this case, $\tw_1$ would be
essentially a wino, and the change in the cross section could extend well
beyond the band in Fig.~\ref{fig:unc}. Fortunately, by studying
chargino pair production we can significantly constrain chargino mixing
angles, or equivalently, the $e\tw_1\tnu_e$ coupling
\cite{fpmt}. We have not studied whether these constraints from
$\tw_1\tw_1$ production are sufficiently restrictive for our purpose.

\section{Can we include other channels?}\label{sec5}

Up to now, we have confined our analysis to just the $\tau\tau jj\ell$ channel
for $\tnu_{\tau}$,
and the analogous $\mu\mu jj\ell$ or $ee jj\ell$ channels for the other
sneutrinos . We remind
the reader that for the first channel, the lepton $\ell$ is defined to
be an electron or muon not coming from the decay of a $\tau$. We saw
that the precision attained is very sensitive to how close we are able
to go to the particle threshold where the signal becomes very small. It
is natural to ask whether we can indeed go closer to the threshold by
including other channels, and thereby improve the precision.

Clearly, we want to confine ourselves to channels with multiple jets and
leptons for which the signal is not obviously overwhelmed by SM
backgrounds or by contamination from other SUSY sources.  We therefore
do not consider direct decays of $\tnu_{\tau}$ to $\tz_1$ since we then
have final states with fewer jets/leptons, and hence SM backgrounds and
SUSY contamination much larger than the signal. For
$\tnu_{\tau}\tnu_{\tau}$ production, the decay chains that we are left
with consist of, \setcounter{equation}{0}
\begin{eqnarray}
\tnu_\tau \tnu_\tau \to \tau \tw_1 \tau \tw_1 && \to \tau \tau j j \ell
+\eslt \\ &&\to\tau \tau j j j j +\eslt \\ &&\to\tau \tau \ell \ell
+\eslt  \\ \tnu_\tau\tnu_\tau \to \nu_\tau \tz_2 \tau \tw_1 &&\to \tau j
j \ell +\eslt \\ &&\to \tau j j j j +\eslt \\ &&\to \tau \ell^+ \ell^-
\ell +\eslt  \\ &&\to
\tau j j \ell^+ \ell^- +\eslt \\ \tnu_\tau \tnu_\tau \to \nu_\tau \tz_2
\nu_\tau \tz_2 &&\to j j j j +\eslt \\ &&\to j j \ell^+ \ell^- +\eslt \\ &&\to \ell^+
\ell^- \ell^+ \ell^- +\eslt \;\;,
\end{eqnarray}
\setcounter{equation}{4}
where we have omitted invisible decays of $\tz_2$ for the same reason.
Muon or electron sneutrino production leads to analogous
decay chains that we do not list here.

Channels 8--10 are strongly contaminated by $\tnu_e\tnu_e$
and $\tnu_{\mu}\tnu_{\mu}$ pair production and also by the production of
lighter charginos and neutralinos, so it seems pointless to consider
these any further.

We have used ISAJET to compute the cross sections after all cuts and
beamstrahlung/ISR corrections in channels 1--7 for both the cases that
we have examined. The contributions to each of these channels from the
main SUSY sources, along with our estimates of the main $2\to 2$ SM
backgrounds\footnote{$WWZ$ production which has a cross section of
4.8~fb is a potential background for some of the event topologies. We do
not list its contribution here as we have not simulated these
events. We expect though that this contribution is either negligible (as
for the $\tau\tau jj\ell$ channel),
substantially smaller than other backgrounds listed (as for the $\tau
jj\ell$ channel), or easily removeable (as for the $\tau \ell\ell\ell$ channel.)}
are shown in Table \ref{tab:tauback} for $\sqrt{s}=500$~GeV. We list
separately the contributions from $\tnu_{\tau}\tnu_{\tau}$ production,
from the production of the first two generations of sneutrinos and
sleptons, from $\ttau_1\ttau_2+\ttau_2\ttau_2$ production (which we
separate out because in models with large $\tan\beta$ this can have a
threshold that is quite different from the threshold for other sleptons
and sneutrinos) and, finally, from the production of chargino and
neutralino pairs. There are two entries for the SUSY contributions; the
first is for Case~I, while the second one (in parenthesis) is for Case~II.

We see that channels 4 and 5 have large backgrounds from $t\bar{t}$
production exceeding the signal by a factor of 29 (26) and 173 (67) for
Case~I (Case~II), respectively.\footnote{Channel 5 also has an
unexpected background of about 1~fb from $W^{\pm}W^{\mp}$ production. We
have traced this to events where $W \to q\bar{q'}gg$. This background
should be reducible by an invariant mass cut on the $4j$ system, but at
some cost to the already small signal.} While vetoing events with $b$
jets together with top reconstruction cuts may well reduce this
background considerably, it will be at some cost to the signal which, in
each channel, starts out at a level of just $\sim 0.1-0.25$~fb (0.29~fb) in Case~I (Case~II). Moreover, the contamination from sleptons
and charginos/neutralinos is several times the signal that cuts designed
to reconstruct the top background will not significantly reduce.

Channel 3 with the $\tau\tau\ell\ell$ final state has a significant
background from $ZZ$ production, though this should be almost eliminated
by an invariant mass cut on the $\ell\ell$ system. The bigger problem is
that this channel is contaminated by events from SUSY sources: the
sleptons, charginos/neutralinos and the staus, each lead to
$\tau\tau\ell\ell$ cross sections that exceed the corresponding signal cross
section by a factor of 6--8, with $\ttau$ events (which
would most resemble the signal) contributing about 1/3 (2/3) of the SUSY
background.

In channels 6 and 7, although the SM contamination is small, the signal
is very tiny. Morover, the SUSY
contamination from the first two generations of sneutrinos and charged
sleptons exceeds the signal by a factor of 25-100.

This leaves us with channels 1 and 2. The decays $Z^0 \to q\bar{q}gg$
result in an unexpected source of SM background in channel 2. The
$t\bar{t}$ background, while smaller, is not insignificant. Contamination
from chargino/neutralino and stau production is comparable to the
signal. While it may well be possible to include this channel at
$\sqrt{s}=500$~GeV, the real problem with this channel comes because we
need the same channels in our energy scan. For instance, at an energy
15~GeV above the threshold, the $Z^0Z^0$ background exceeds the signal
by a factor 40 (15) for Case~I (Case~II), while the top background
(present only for Case~II) exceeds the signal by a factor $\sim
6$. Considering all these problems together with the fact that the
signal in Channel 2 would add just 30-50\% to Channel 1, we did not
think that it was worth including it in our analysis in Sec.~\ref{sec4}.

We have also examined whether it might be possible to include other
signals from $\tnu_{\mu}\tnu_{\mu}$ to obtain a significantly better
determination of $m(\tnu_{\mu})$. Toward this end, we have computed the
cross section in the appropriate channels analogous to those shown in
Table \ref{tab:tauback}, along with the SM backgrounds. Our results are
shown in Table \ref{tab:muback} for $\sqrt{s}=500$~GeV. Again the
entries in parenthesis refer to the cross sections in Case~II. Here, in
the last two rows of both the SUSY sources as well as the SM
backgrounds, to avoid double counting (with channels 3 and 1), the
$\ell$ in the case of the $\mu\ell\ell\ell$ and $\mu jj \ell\ell$ channels
refers only to any $e$ or a $\mu$ from the decay of a tau (which, we
have assumed, can be tagged by a vertex detector). We should mention
that, in this case, for $\ell=\mu$ channel 4 would be the same as
channel 9. A similar comment applies {\it vis \`a \ vis} channels 3 and
10. Unfortunately, essentially the same reasoning as before leads us to
conclude that it is not possible to improve the sneutrino mass
measurement by including other channels.\footnote{The largest signal is
in channel 4. Unfortunately, even if the large top background can be
controlled by $b$-veto and top reconstruction, contamination from other
SUSY sources is still large. It may be interesting to see whether it is
possible to devise cuts to reduce this background without killing the
signal.}

Finally, we turn to the case of electron sneutrinos. As we noted in the
last section, the question in this case is whether we can include other
channels close to the $\tnu_e\tnu_e$ threshold so as to be able to
cleanly extract the threshold behaviour of the signal cross section.
The dominant SUSY contamination at an energy about 2 GeV above the
nominal electron sneutrino threshold which mainly comes from the
production of charginos and neutralinos is shown in Table
\ref{tab:eback} along with the $\tnu_e$ signal in various channels. In
addition, SM backgrounds from $W$ and $Z$ pair production, as well as
$t\bar{t}$ production (only for Case~II) will also be present. It should
be remembered that the cross section for $WW$ production is now very
large because $P_L=0.9$. Backgrounds from $t\bar{t}$ and $ZZ$ production
are about 1.5-2 times larger for left polarized electron beams
\cite{munroe}. Channel 5 will have a large background from $WW$
production (see Table \ref{tab:muback}). For Case~II only, $t\bar{t}$
production will be a formidable SM background in channels 4 and 5. In
channels 6 and 7, the signal is very small. This leaves us with channels
1-3.  Neutralino production (mainly $\tz_2\tz_2$) contaminates channel
3, particularly in Case~II for which it has considerable phase space
even at the sneutrino threshold.
It is possible, of course, that the signal from
sneutrinos may be relatively enhanced by further cuts, {\it e.g.} on
$m(\ell^+\ell^-)$ or $p_T(e)$, but such a detailed analysis is beyond
the scope of this study.  Finally, we remark that the signal in the
remaining (relatively background-free) channel 2 may be combined with
that in channel 1 resulting in an increase of about 40\%. Presumably,
this will increase the precision from that shown in
Fig~\ref{fig:eellipse} by about 20\%.

We should add that in favourable cases, considerably better precision
might be possible. For instance, if $t\bar{t}$ production is
kinematically forbidden close to the threshold (as in Case~I), it should
be possible to combine the signal in channel 4 with those in channels 1
and 2, resulting in a three-fold increase in the signal. Optimistically,
one may even assume that this may be possible even in Case~II since top
reconstruction and $b$-veto may severely reduce the top
background. Nonetheless, one sees that even if we combine the signals in
all the channels the signal is increased by at most a factor $\sim
5$. If we further assume that both SM and SUSY backgrounds can be
eliminated without loss of signal, this increased signal is
statistically equivalent to an increase in available integrated
luminosity. We conclude that even in such a favourable scenario, the
precision on $m(\tnu_e)$, for an integrated luminosity of $\sim
100$~fb$^{-1}$ will be comparable to that given by the inner ellipse in
Fig.~\ref{fig:eellipse}, but still much larger than the 70~MeV
claimed in Ref. \cite{tesla}. In this case width effects, that we have
neglected here, would become
important for an analysis of what might be achievable with a data sample
of 500~fb$^{-1}$.

\section{Comments and Conclusions}\label{sec6}

We have performed a detailed examination of the prospects for the
determination of sneutrino masses at future linear colliders via a
measurement of the energy dependence of the cross section. Threshold
studies, which have been claimed\cite{tesla} to yield a precision better
than parts per mille (0.5\%) with an integrated luminosity of just
100~fb$^{-1}$ for $m(\tnu_e)$ ($m(\tnu_{\mu})$ and $m(\tnu_{\tau}$)), are a
special case of this.

We find that for the second and third generations, the sneutrino
production rate leads to less than 1 event for each energy point if we
adopt the energy scan strategy and luminosity proposed in
Ref. \cite{tesla} and restrict ourselves to the $\mu\mu jj\ell$ and
$\tau\tau jj\ell$ channels that have been suggested to be relatively
free of SM backgrounds and SUSY contamination. The availability of 80\%
positron beam polarization increases the signal by about 40\%, but this
is not sufficient. Also, including other channels did not appear to help
significantly, since in those channels where the signal was substantial,
SUSY contamination is also large.

We have also identified a different issue that degrades the precision
with which sneutrino masses can be determined from counting experiments,
even if the signal is large. The point is that the size of the signal
depends on an unknown branching fraction for sneutrinos to decay to the
channel of interest. For sneutrinos, this is a problem of principle
because
even if backgrounds are small and even with a
perfect detector, sneutrinos decaying via $\tnu
\to \nu\tz_1$ (whose branching fraction can be anything up to 100\%)
always escape detection. We emphasize that although we are discussing
this only in the context of sneutrinos, this issue arises even for
charged sparticle mass measurements (except, possibly, for the lightest
charged particle as it can only decay to the LSP) in realistic detectors:
although SM backgrounds are generally greatly reduced by suitable cuts,
to eliminate SUSY contamination, one is frequently forced to study the
signal in specific channels whose branching fraction is again to be
determined using the same data that is used to extract the mass.  Thus,
the sparticle mass and this branching fraction have to be simultaneously
extracted by fitting both these to the same data. This considerably
degrades the precision with which the mass can be determined.  While
this is, in principle, also a problem for the extraction of masses using
kinematic strategies \cite{jlc,mur,munroe,zdr,snownlc}, the sensitivity
to the unknown branching fraction is presumably smaller.

For sparticles whose cross sections into channels where SUSY
contamination and SM backgrounds are under control is relatively small,
we found that the following offers a reasonable strategy for extracting
their mass:
\begin{itemize}
\item Use about 15-20\% of the available integrated luminosity at the
nominal collider energy which, we assume is well beyond the sneutrino
production threshold. In our analysis, we took this to be 500~GeV. A
measurement of the cross section at this energy strongly constrains the
branching fraction.

\item Divide the remaining integrated luminosity to measure the cross
section at about three additional lower energy points in such a manner
so as to obtain about an equal number of events at each point so that
the fractional error in the cross section at each energy is about the
same. Since the precision on the mass improves if we are able to go
close to the threshold where the cross section is small and a large
integrated luminosity is required to get a fixed number of events, we
found that it was best to space these points $\Delta \sim 60$~GeV
apart. This is because two of the three points are away from the
threshold and so can reach the target number of events with a modest
integrated luminosity.

\end{itemize}

Specifically, for muon and tau type sneutrinos, the availability of
right-handed electron beams was essential to reduce the contamination
from electron sneutrinos and chargino production.  We made several
optimistic assumptions for our analysis. First, we assumed that 95\%
electron beam polarization will be available. Second, for the
determination of $m(\tnu_{\tau})$, we assumed that vertex detection
would allow leptonically decaying taus to be tagged with an efficiency
close to 100\%. While this would also allow hadronically decaying taus
to be discriminated from light quark and gluon jets, the need to
discriminate tau jets from $c$-jets required us to impose requirements
on hadronically decaying taus that reduce their efficiency, but maintain
the purity of the tau sample. We should, however, remind the
reader that we have used the beamstrahlung spectrum appropriate to
$\sqrt{s}=500$~GeV all the way down to the sneutrino mass
threshold. This will cause us to somewhat underestimate
the precision that might be attained.

Even so, we found that an integrated luminosity of 100~fb$^{-1}$ was too
small for a measurement of $m(\tnu_{\tau})$. This is because we were
forced to confine our analysis to the cleanest $\tau\tau jj\ell$
channel. Other channels either had small cross sections or suffered from
considerable SUSY backgrounds.  A crude determination of $m(\tnu_{\mu})$
may be possible\footnote{Fig.~\ref{fig:muback} allows the reader to
assess how small $D$ can be to maintain the six event minimum as we
have required.} at least in Case~II even with an integrated luminosity
of $\sim 100$~fb$^{-1}$, but since $m(\tnu_{\mu})$ and $m(\tnu_{\tau})$
can be determined via essentially the same collider runs, we have
assumed an integrated luminosity of 500~fb$^{-1}$ for both
analyses. Even then, we found that, at best, a measurement of these
masses with a precision of several percent (at 90\% CL) is possible,
with of course $m(\tnu_{\mu})$ being $2-3$ times better determined than
$m(\tnu_{\tau})$. The exact precision that can be attained depends on
the branching fraction into the useful channels which can vary by a
factor $2-3$ for representative ranges of SUSY parameters.\footnote{Of
course, there are regions of parameter space where the variation may be
outside this range.} The attainable precision will also depend to some
extent on how well SUSY backgrounds may be controlled. The main results
of our studies are shown in Fig.~\ref{fig:tauellipse} for $\tnu_{\tau}$
and Fig.~\ref{fig:muellipse} for $\tnu_\mu$, and summarized in
Table~\ref{tab:summary}. We emphasize again that these results are for
a model-independent analysis. As discussed in Sec.~\ref{sec4}C, analyses
within particular frameworks such as mSUGRA may yield much higher precision.

It also seems worthwhile to stress that unlike a closely spaced
threshold scan the strategy of taking data at widely spaced intervals
would conceivably be useful for other physics, than just sneutrino
masses. For instance, the same energy scan could be used to
simultaneously obtain masses of several sparticles or even make
measurements useful for Higgs boson, top as well as electroweak or QCD
studies.

We also examined prospects for determining $m(\tnu_e)$. For this, of
course, left-handed electron polarization is optimal. Again, it appeared
to be best to use about 20\% of the available luminosity at the nominal
energy, and divide the remainder among about three points at lower
energies with equal number of events at each point. In contrast to the
second and third generation cases, it seemed best to choose the points
just about 1~GeV apart, though the precision for a 30~GeV spacing is not
much worse as can be seen in the first row of
Fig.~\ref{fig:eellipse}. Indeed the consideration in the previous
paragraph may suggest that this may be better for the overall physics 
program at the LC. The reader should keep in mind that even with this strategy
most of the data will be at the one point closest to the threshold. With
an integrated luminosity of 120~fb$^{-1}$, a precision of about 0.5\%
(90\% CL) appears to be possible.  This will improve by a factor 2 for a
data sample of 500~fb$^{-1}$. In this case, some improvement may be
possible by combining several channels.
The dotted ellipses in the second row show
that the precision on $m(\tnu_e)$ will be degraded by about 25-30\% if,
in the interest of the overall LC program, it is decided to collect
equal amounts of integrated luminosity at points spaced 30~GeV apart.

To conclude, we reiterate that experiments at the LC will be able to
determine masses of sneutrinos with significant precision but certainly
for the third
generation, and possibly also second generation, sneutrino mass
determination an integrated luminosity $\sim 500$~fb$^{-1}$ seems
essential. However, even with this large luminosity, the precision
claimed in Ref.\cite{tesla} does not seem possible. While we have not
checked this, the mass resolution for the heavier charged sleptons
$\tmu_L$ and $\ttau_2$ is presumably not any better than that for the
corresponding sneutrinos.  The masses of first generation sneutrinos can
be determined with much greater precision, but even in this case, our
projections for the precision are nowhere near as optimistic as those in
Ref. \cite{tesla}. We caution that several beautiful analyses
\cite{tesla,porod,kalinowski} about what might be attainable in light of
measurements at a LC might be painting too rosy a picture.

\acknowledgements We thank G. Blair and H.-U. Martyn for communications
at the early stages of this study. We are grateful to R.M.~Godbole for
her continuing interest, and for comments on an early draft of this
paper.  This research was supported in part by the U.~S. Department of
Energy under contract number DE-FG02-97ER41022 and DE-FG-03-94ER40833.

%
\begin{table}
\begin{center}
\caption{Relevant sparticle masses in GeV and branching fractions for Case~I
and Case~II introduced in Sec.~\ref{sec1} of the text. The last line
shows the branching fraction into the relatively background free
$\tau\tau jj\ell$ channel discussed in Sec.~\ref{sec2}.}
\bigskip
\begin{tabular}{lcc}
\hline
 & Case~I & Case~II  \\
\hline

$m(\te_R)$ & 130.4   & 167.8 \\
$m(\te_L)$ & 173.7   & 194.2 \\
$m(\tnu_e)$ & 158.3 & 178.3 \\
$m(\ttau_1)$ & 129.3  & 165.7  \\
$m(\ttau_2)$ & 174.4 & 195.4  \\
$m(\tnu_{\tau})$ & 158.3 & 178.1  \\
$m(\tw_1)$ & 128.9 & 105.6  \\
$m(\tw_2)$ & 340.2 & 283.4 \\
$m(\tz_1)$ & 72.0 & 59.9  \\
$m(\tz_2)$ & 131.4 & 108.2  \\
$m(\tz_3)$ & 313.5 & 255.5  \\
$m(\tz_4)$ & 343.0 & 284.3  \\
$m_h$ & 99.4 & 105.0  \\
$m_A$ & 366.1 & 310.5  \\
$\mu$ & 309.7 & 248.4\\
\hline
$B(\tnu_{\tau} \to \tau\tw_1)$ & 0.37 & 0.56 \\
$B(\tw_1 \to \ell\nu\tz_1)$ & 0.154 & 0.124 \\
$B(\tw_1 \to q\bar{q}\tz_1)$ & 0.523 & 0.626\\
$B(\tnu_{\tau}\tnu_{\tau} \to \tau\tau jj\ell)$ & 0.044 & 0.097
\label{tab:cases}
\end{tabular}
\end{center}
\end{table}

\begin{table}
\begin{center}
\caption{The cross section after cuts for the $\tau\tau jj\ell$ signal
at $\sqrt{s} = 500$ GeV for several channels for the two cases discussed
in the text. Here $j^{\tau}$ and $\ell^{\tau}$ refer to a jet or lepton
($e$ or $\mu$) from the decay of a tau, which itself is produced via the
decay of a heavy particle. The parameter $r_{min}$ which is used to
discriminate between $\tau$ and $c$ jets has been introduced in
Sec.~\ref{sec4}. We see that if the taus can be tagged via
their leptonic decay, the signal is more than doubled.}
\bigskip
\begin{tabular}{ccccc}
topology &  $r_{min}$  & $\sigma^{obs}$ (fb) (Case I) & $\sigma^{obs}$ (fb)
(Case II)\\
\hline
                                       & 0.5  & 0.045  &  0.118 \\
$2 j^\tau + 2 j^{QCD} + \ell + \eslt$  & 0.8  & 0.040  &  0.114 \\
                                       & 1.0  & 0.028  &  0.094 \\
\hline
                                       & 0.5  & 0.051  &  0.130 \\
$j^\tau + \ell^\tau + 2 j^{QCD} + \ell + \eslt$  & 0.8 & 0.047 &  0.129 \\
                                       & 1.0  &  0.039 &  0.119  \\
\hline
$\ell^\tau + \ell^\tau + 2 j^{QCD} + \ell + \eslt$  & -- & 0.015     &  0.033

\label{tab:tau}
\end{tabular}
\end{center}
\end{table}

\renewcommand{\baselinestretch}{1.2}

\begin{table}[htb]
\begin{center}
\caption{The cross section in fb after all the cuts but no beamstrahlung/ISR
corrections for channels 1--7 discussed in Sec. \ref{sec5} of the text
from several SUSY sources along with SM backgrounds at $\sqrt{s} = 500$
GeV. The first of the SUSY contributions is for Case~I while the second
one (in parenthesis) is for Case~II.  }
\bigskip
\begin{tabular}{ccccc}
  & \multicolumn{4}{c } {SUSY SOURCES} \\
\hline
Channel   & $\tnu_\tau \tnu_\tau$ & $\tl_L, \tnu_\ell$ & $\tw_i, \tz_j$  &
$\ttau_1, \ttau_2$  \\
\hline
$\tau \tau j j \ell$   &
0.110 (0.281)  &  0.049 (0.182) & 0.054 (0.159)  & 0.024 (0.027)\\
$\tau \tau j j j j$ &
0.033 (0.144)  &  0.000 (0.000) & 0.030 (0.145)  & 0.002 (0.091)\\
$\tau \tau \ell \ell$  &
0.088 (0.164)  &  0.363 (0.430) & 0.080 (0.224)  & 0.240 (0.622)\\
\hline
$\tau j j \ell$        &
0.258 (0.291)  &  1.24 (1.34) &  1.08 (0.774) & 0.463 (0.296) \\
$\tau j j j j$         &
0.113 (0.293)  &  0.005 (0.001) &  0.616 (1.15) & 0.040 (0.379) \\
$\tau \ell \ell \ell$  &
0.012 (0.049)  &  1.50 (1.39) &  0.099 (0.145) & 0.074 (0.092)\\
$\tau j j \ell \ell$   &
0.007 (0.043)  &  0.667 (1.25) &  0.069 (0.184) & 0.045 (0.092) \\
\hline
\hline
   &    &    &   & \\
  & \multicolumn{3}{c } {SM BACKGROUNDS} \\
\hline
Channel  & $W^\pm W^\mp$      & $Z^0 Z^0$ & $t \bar{t}$  \\
\hline
$\tau \tau j j \ell$   &  0.000  & 0.002 &  0.003  \\
$\tau \tau j j j j$    &  0.000  & 0.098 &  0.068  \\
$\tau \tau \ell \ell$  &  0.000  & 0.557 &  0.000 \\
\hline
$\tau j j \ell$        & 0.006  & 0.001 &  7.57  \\
$\tau j j j j$         & 1.17  & 0.063 &  19.5  \\
$\tau \ell \ell \ell$  & 0.000  & 0.000 &  0.000  \\
$\tau j j \ell \ell$   & 0.000  & 0.000 &  0.011
\label{tab:tauback}
\end{tabular}
\end{center}
\end{table}

\begin{table}[htb]
\begin{center}
\caption{The same as Table~\ref{tab:tauback} but  for channels with potential
signals from $\tnu_\mu$ production. For the $\mu\ell\ell\ell$ and $\mu
jj\ell\ell$ channels, the $\ell$ refers to any electron or a muon from
tau decay. This avoids double counting as discussed in the
Sec.~\ref{sec5} of the text.}
\bigskip
\begin{tabular}{cccc}
  & \multicolumn{3}{c } {SUSY SOURCES} \\
\hline
Channel   & $\tnu_\mu \tnu_\mu$ & $\tl_L, \tnu_\ell$ & $\tw_i, \tz_j$ \\
\hline
$\mu \mu j j \ell$   &
0.288 (0.539) & 0.177 (0.451) & 0.118 (0.276)  \\
$\mu \mu j j j j$ &
0.084 (0.244) & 0.002 (0.149) & 0.050 (0.225)  \\
$\mu \mu \ell \ell$  &
0.264 (0.341) & 1.21 (1.16) & 0.139 (0.380)  \\
\hline
$\mu j j \ell$        &
0.506 (1.06) & 2.45 (2.89) & 1.58 (2.01)  \\
$\mu j j j j$         &
0.113 (0.298) & 0.106 (0.501) & 0.927 (1.54)  \\
$\mu \ell \ell \ell$  &
0.009 (0.020) & 1.07 (1.20) & 0.047 (0.070)  \\
$\mu j j \ell \ell$   &
0.007 (0.031) & 0.822 (1.74) & 0.057 (0.141)  \\
\hline
\hline
    &    &   & \\
  & \multicolumn{3}{c } {SM BACKGROUNDS} \\
\hline
Channel  & $W^\pm W^\mp$      & $Z^0 Z^0$ & $t \bar{t}$  \\
\hline
$\mu \mu j j \ell$   &  0.000  & 0.000  &  0.003  \\
$\mu \mu j j j j$    &  0.000  & 0.001 &  0.122  \\
$\mu \mu \ell \ell$  &  0.000  & 0.039 &  0.000 \\
\hline
$\mu j j \ell$        & 0.001  & 0.317 &  8.17  \\
$\mu j j j j$         & 1.35   & 0.001 &  23.0  \\
$\mu \ell \ell \ell$  & 0.000  & 0.000 &  0.000  \\
$\mu j j \ell \ell$   & 0.000  & 0.001 &  0.000
\label{tab:muback}
\end{tabular}
\end{center}
\end{table}

\begin{table}[htb]
\begin{center}
\caption{The cross section in fb for potential signals from
$\tnu_e\tnu_e$ production at $\sqrt{s}=320$~GeV (360~GeV) along with
chargino and neutralino contamination for Case~I (Case~II). In this
table, $P_L= 0.9$. For the $e\ell\ell\ell$ and $e jj\ell\ell$ channels,
the $\ell$ refers to any muon or a electron from tau decay. This is to
avoid double counting.}

\bigskip
\begin{tabular}{ccc}
Channel   & $\tnu_e \tnu_e$  & $\tw_i, \tz_j$ \\
\hline
$e e j j \ell$   &
0.233 (0.533) & 0 (0)   \\
$e e  j j j j$ &
0.080 (0.207) & 0 (0)  \\
$e e \ell \ell$  &
0.247 (0.420) & 0.225 (1.68)   \\
\hline
$e j j \ell$        &
0.514 (1.45) & 0.188 (6.59)   \\
$e j j j j$         &
0.111 (0.276) & 0.013 (0.040)   \\
$e \ell \ell \ell$  &
0.004 (0.023) & 0 (0)   \\
$e j j \ell \ell$   &
0.004 (0.027) & 0 (0)   \\
\hline
\label{tab:eback}
\end{tabular}
\end{center}
\end{table}

\begin{table}[htb]
\begin{center}
\caption{A summary of our projections for sneutrino mass measurements
(90\% CL) in Case~I and Case~II, assuming a 95\% longitudinally
polarized beam. For especially $\tnu_{\tau}$, a significant mass
measurement does not appear to be possible with 100~fb$^{-1}$. For each of
$m(\tnu_{\tau})$ and $m(\tnu_{\mu})$, the first row shows our
projection with backgrounds and SUSY contamination as discussed in the
text, while the next one shows the corresponding projection if these
backgrounds can be effectively eliminated without loss of signal. For
$\tnu_e$, both SM background and SUSY contamination are insignificant.   }

\bigskip
\begin{tabular}{ccc}
  & Case~I  & Case~II \\
\hline

$m(\tnu_{\tau})$ (500~fb$^{-1})$   &
$153^{+12.5}_{-24}$~GeV & $174.9^{+7.1}_{-15.4}$~GeV   \\
   & $153^{+11.5}_{-24}$~GeV & $175.4^{+5.6}_{-10.9}$~GeV   \\
$m(\tnu_{\mu})$ (500~fb$^{-1})$ &
$156.4^{+4.2}_{-7.9}$~GeV & $175.7^{+3.9}_{-4.6}$~GeV  \\
   & $156.4^{+4.1}_{-7.4}$~GeV & $176.4^{+3.3}_{-5.2}$~GeV  \\
$m(\tnu_e)$ (120~fb$^{-1}$) &
$157.8^{+0.8}_{-1.2}$~GeV &$178.0^{+0.5}_{-0.8}$~GeV       \\
$m(\tnu_e)$ (500~fb$^{-1}$) &
$158.1^{+0.4}_{-0.5}$~GeV  & $178.2^{+0.2}_{-0.4}$~GeV
\label{tab:summary}
\end{tabular}
\end{center}
\end{table}

%
\iftightenlines\else\newpage\fi
\iftightenlines\global\firstfigfalse\fi
\def\dofig#1#2{\epsfxsize=#1\centerline{\epsfbox{#2}}}
\def\fig#1#2{\epsfxsize=#1{\epsfbox{#2}}}


\begin{figure}
\noindent
\dofig{17cm}{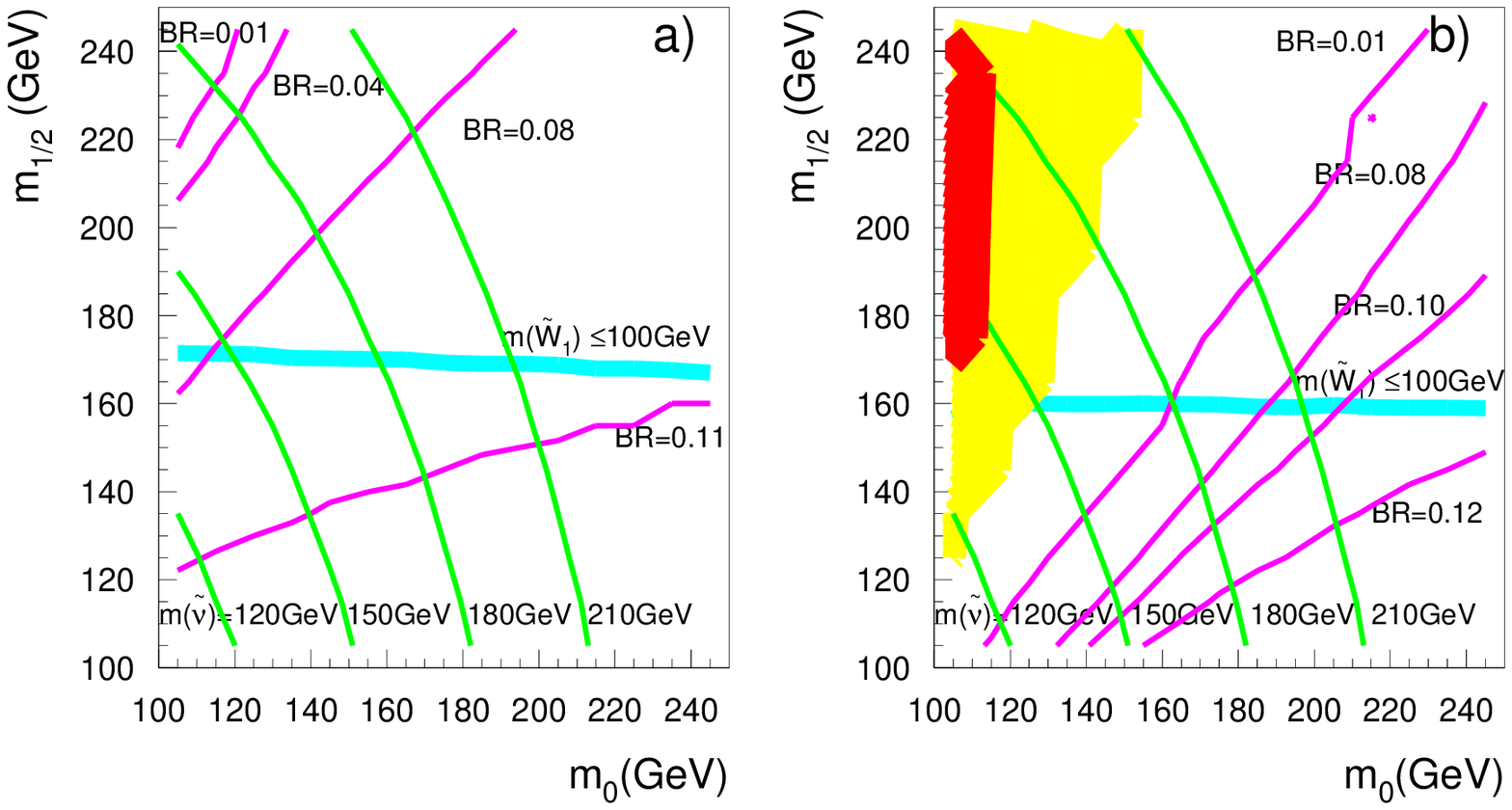}
\hspace*{1cm}
\caption[]{Contours of the branching ratio $\tnu_\tau \tnu_\tau \to \tau
\tw_1 \tau \tw_1 \to \tau \tau j j + \ell + \eslt$ for {\it a})
$\tan\beta =3$ and {\it b}) $\tan\beta =40$. Also shown are contours of
$m(\tnu_{\tau})$. In the region below the thick line, $m(\tw_1) \leq 100$~GeV, roughly
the lower limit on the chargino mass from LEP experiments. The dark
(light) shaded
regions in frame {\it b}) are excluded because $m^2(\ttau_R) < 0$
($\ttau_1$ is the LSP).}
\label{fig:bf}
\end{figure}

\newpage


\begin{figure}
\noindent
\fig{8.5cm}{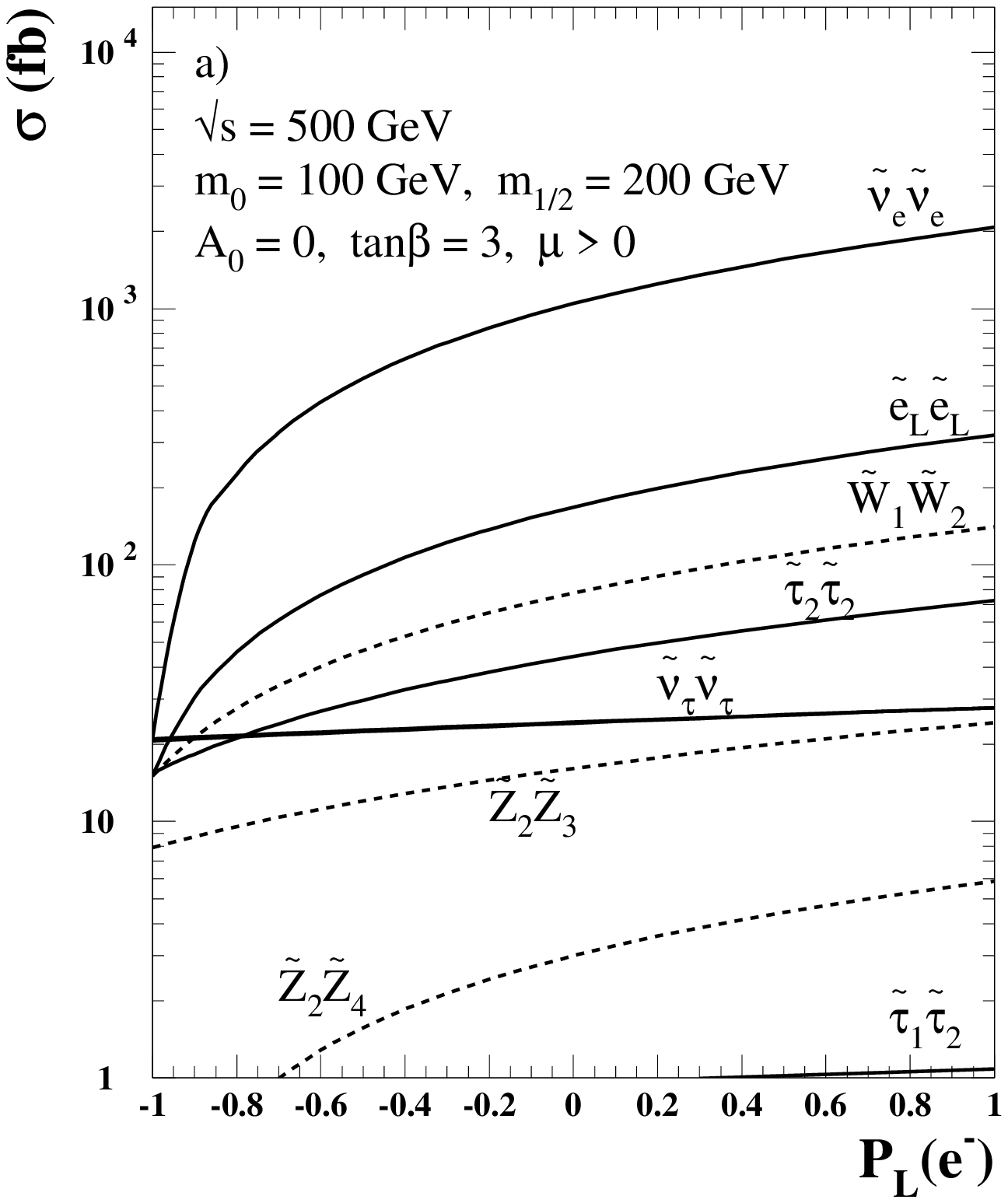} \hspace*{-1.5cm} \fig{8.5cm}{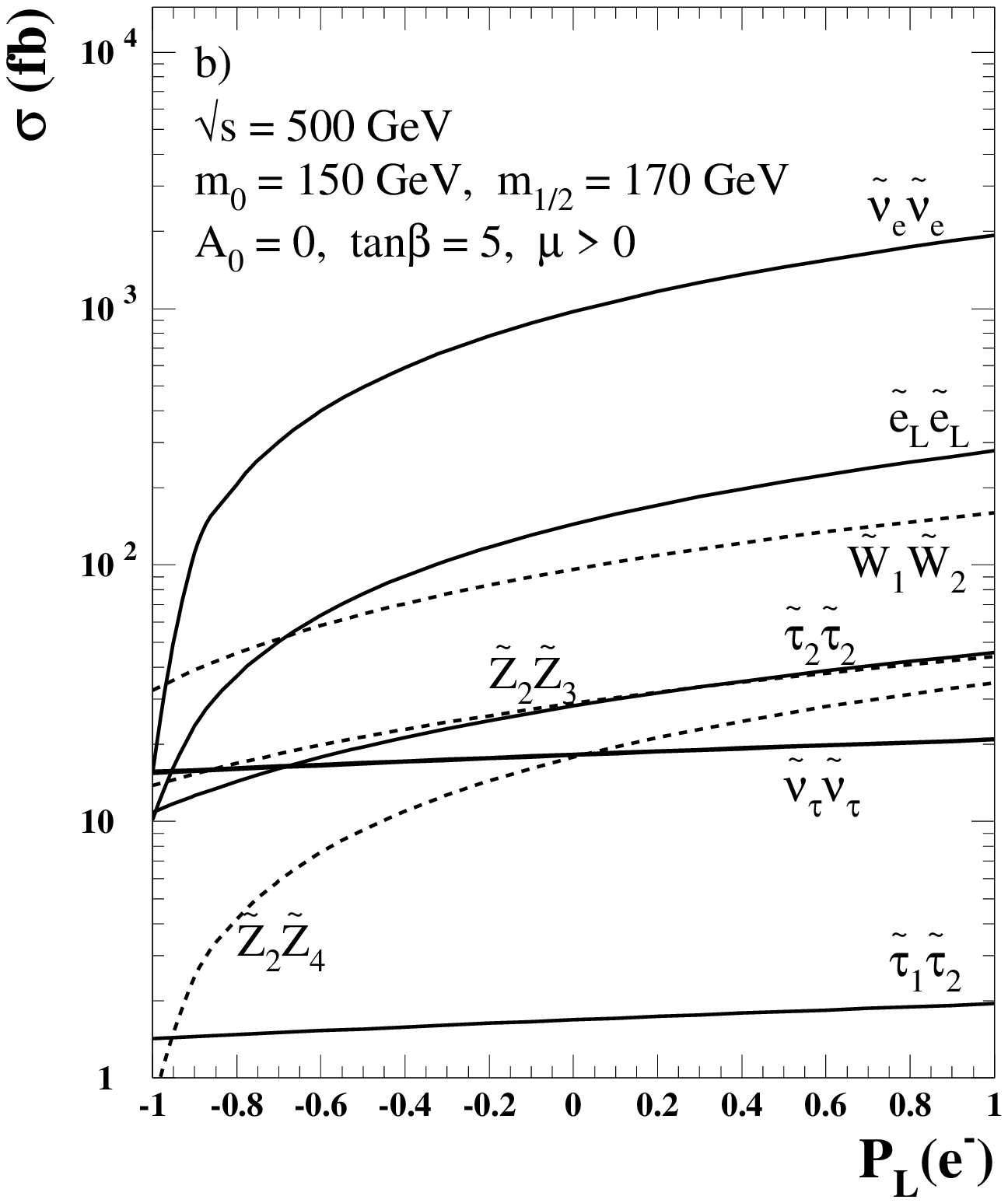}\\
\hspace*{1cm}
\caption[]{Production cross sections for several SUSY processes at a
$\sqrt{s} = 500$ GeV $e^+e^-$ collider versus the electron beam
polarization parameter $P_L(e^-)$. The solid (dashed) lines are the
cross sections for various slepton and sneutrino (chargino and
neutralino) pair production processes. Frame {\it a}) is for Case~I and
frame {\it b}) is for Case~II.}
\label{fig:csections}
\end{figure}

\newpage


\begin{figure}
\dofig{6in}{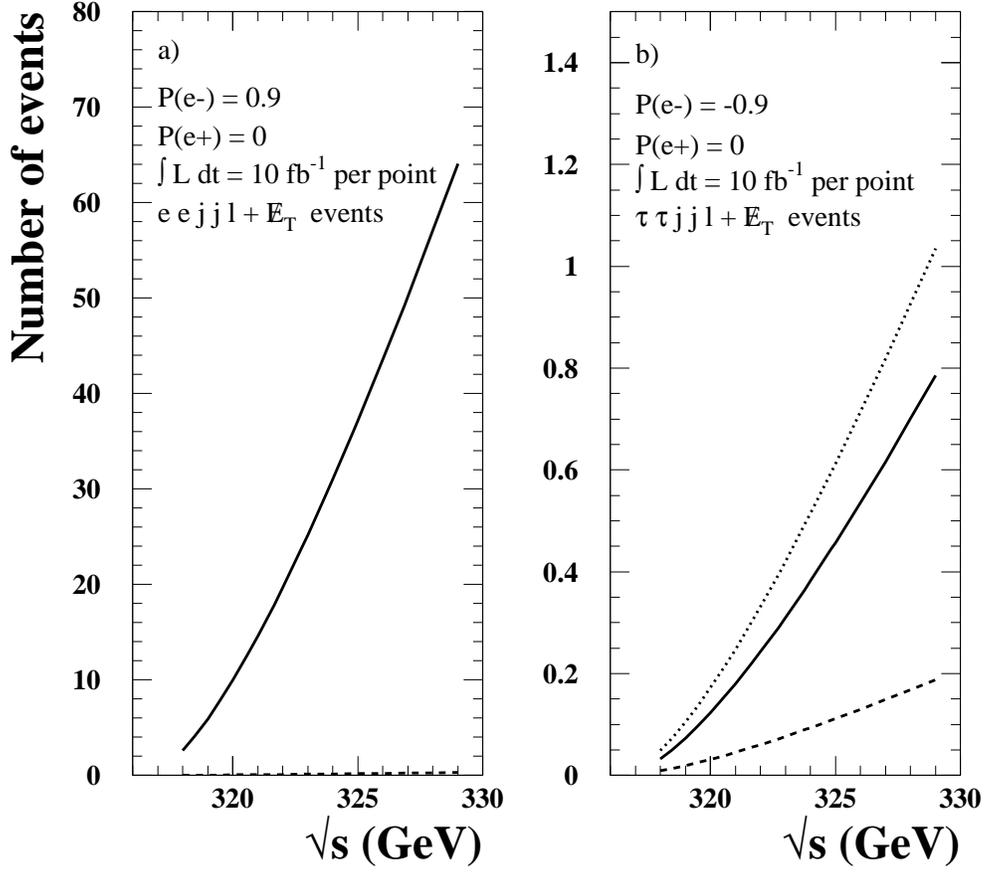}
\caption[]{An illustration of event rates close to the sneutrino
production threshold for Case I. In frame~{\it a}) we show the rate in
the favoured $e e j j \ell$ channel from $\tnu_e\tnu_e$ production with
$P_L(e^-) = 0.9$. The dashed line shows the
contamination from $\tnu_{\mu}\tnu_{\mu}$ production. In frame {\it b}),
we show rates for the corresponding $\tau \tau j j \ell$ channel from
$\tnu_{\tau}\tnu_{\tau}$ production, but with $P_L(e^-)=-0.9$ (dotted
line) along with contamination from $\tnu_e\tnu_e$ production (solid
line) and $\tnu_{\mu}\tnu_{\mu}$ production (dashed line). In both
frames we have assumed that the integrated luminosity is 10~fb$^{-1}$
for each energy point.}
\label{fig:thresh}
\end{figure}


\begin{figure}
\noindent
\hspace*{1cm}
\fig{8cm}{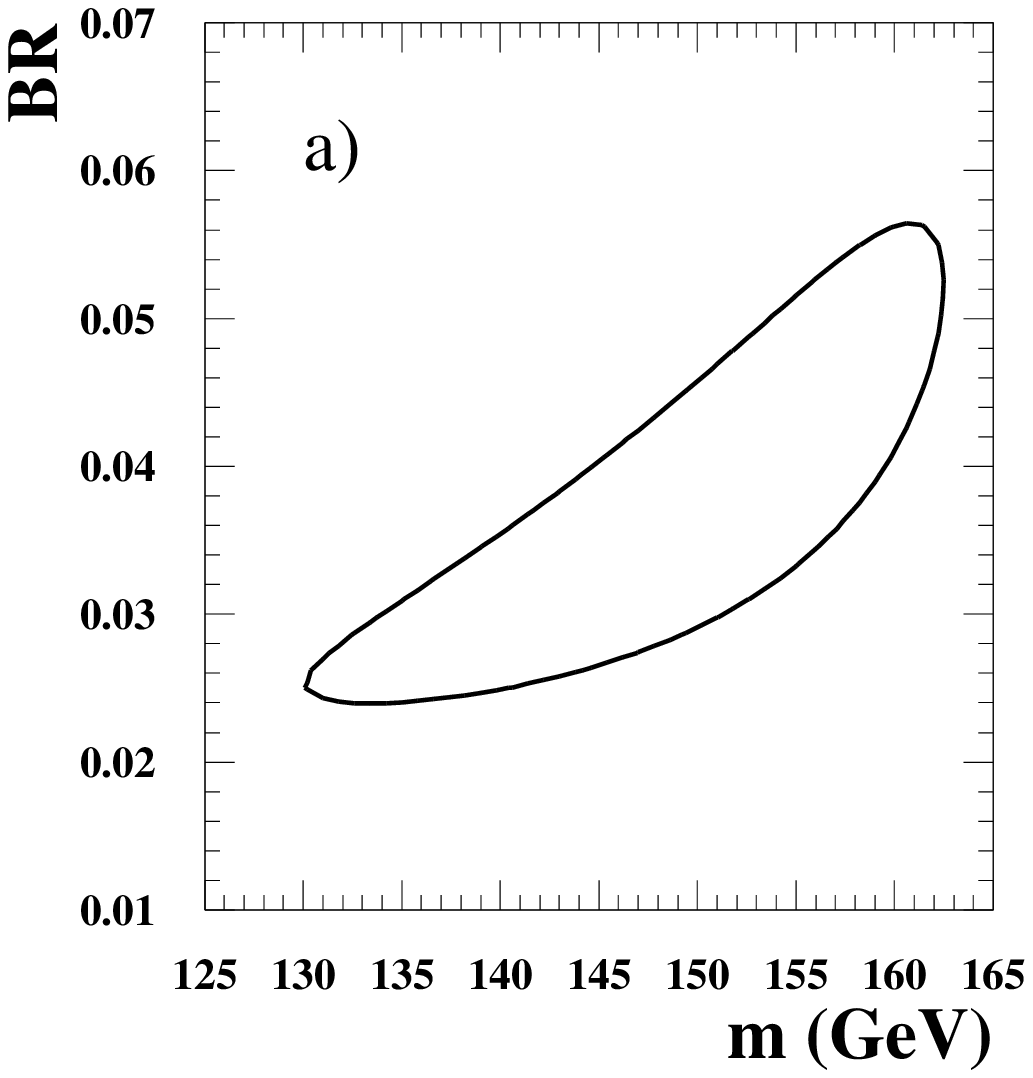} \hspace*{-1cm} \fig{8cm}{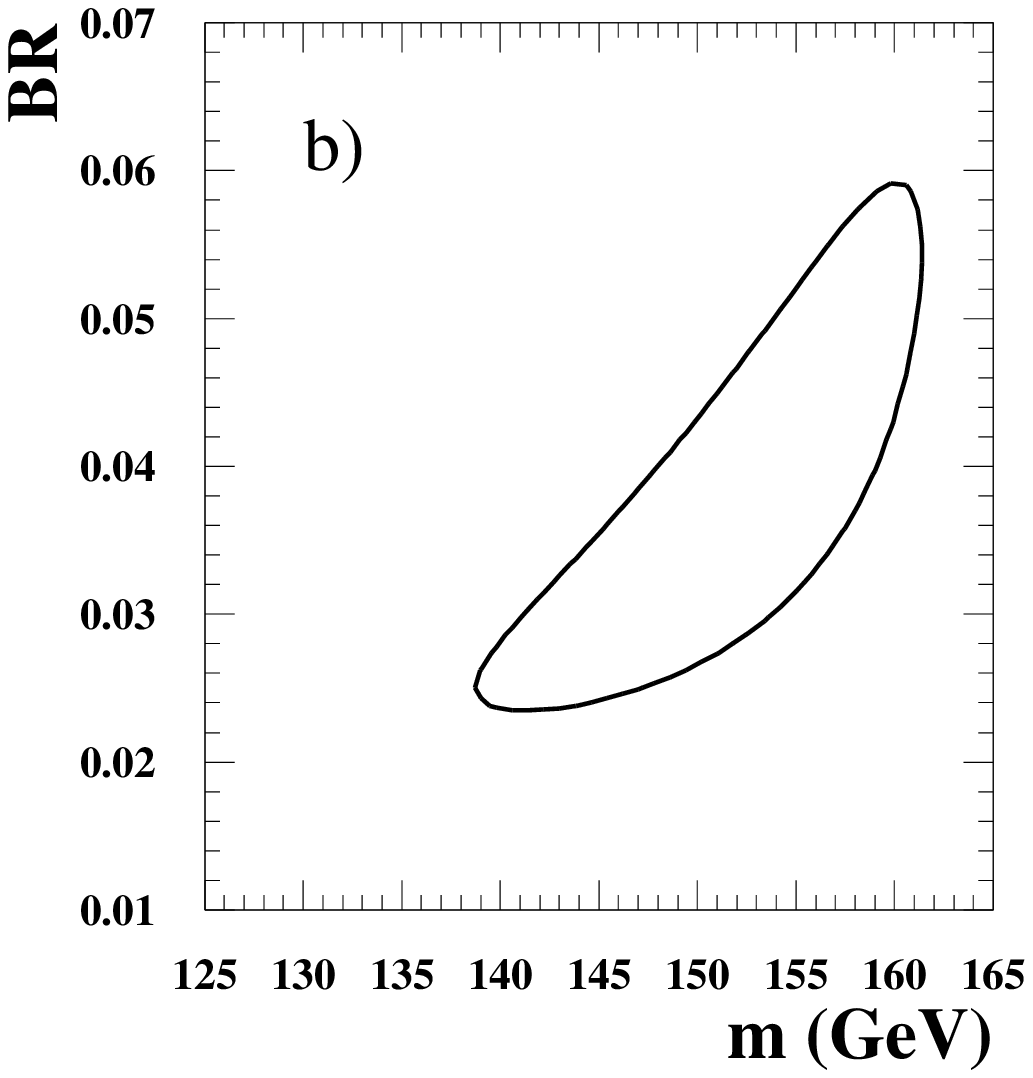}
\vspace*{-1cm}
\\
\hspace*{1cm}
\fig{8cm}{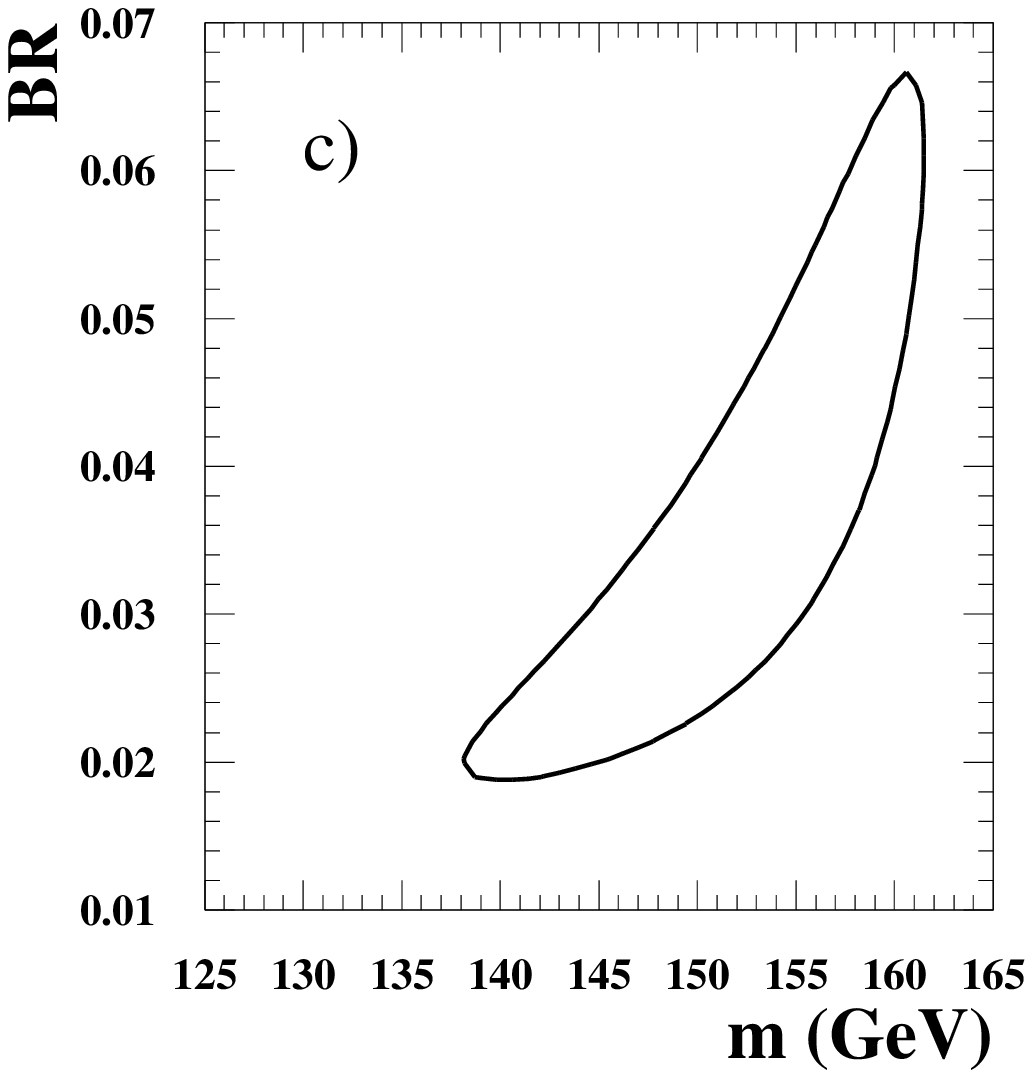} \hspace*{-1cm} \fig{8cm}{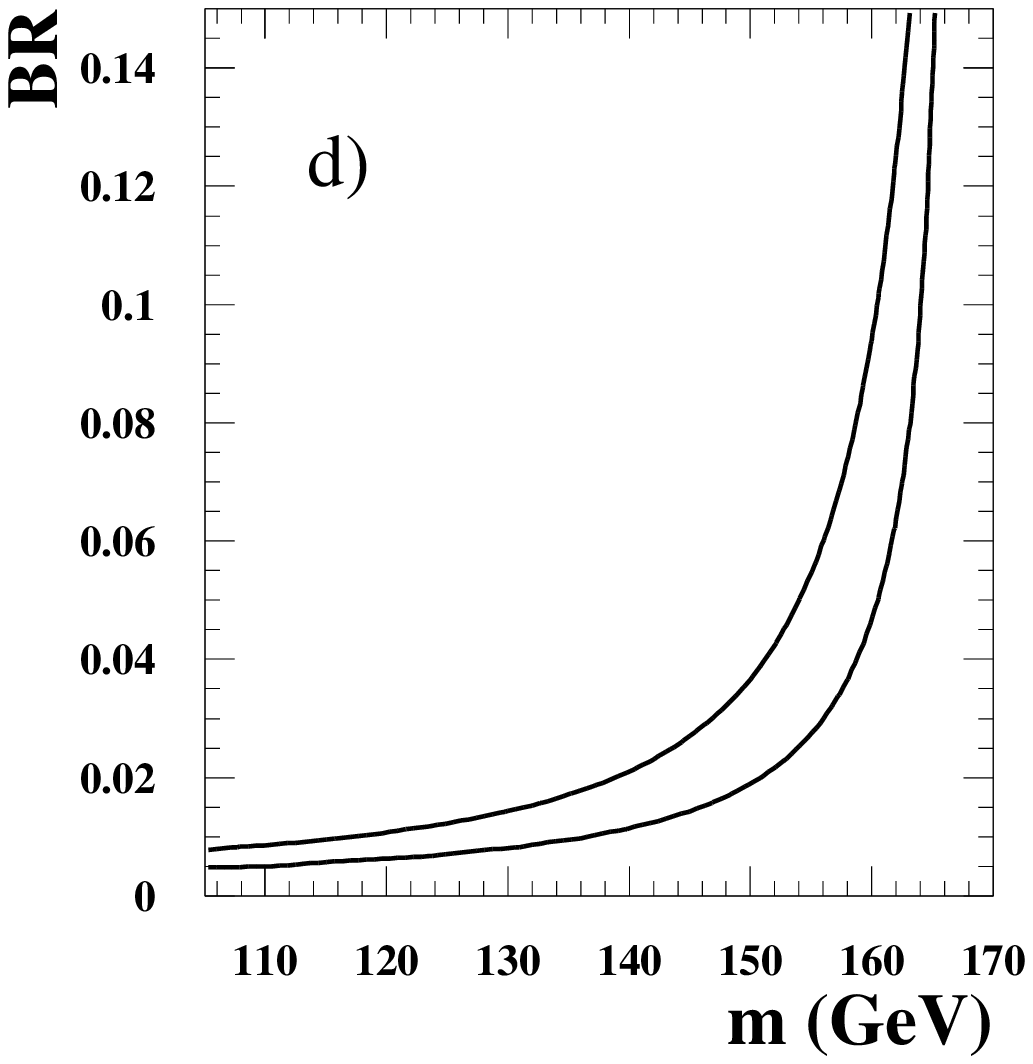}
\vspace*{0.5cm}
\caption[]{90\% CL ($\Delta\chi^2=4.6$) contours for a fixed 10 events per
point (for the input theory) at lower energies, with $N =$ 3 and $\Delta
= 5$ GeV.  In a) $ {\cal L}_{low} = 60 \; \mbox{fb}^{-1}, {\cal L}_0 =
60 \; \mbox{fb}^{-1}$ and $D=$ 45 GeV; in b) ${\cal L}_{low} = 80 \;
\mbox{fb}^{-1}, {\cal L}_0 = 40 \; \mbox{fb}^{-1}$ and $D=$ 32 GeV; in
c) ${\cal L}_{low} = 100 \; \mbox{fb}^{-1}, {\cal L}_0 = 20 \;
\mbox{fb}^{-1}$ and $D=$ 26 GeV, and in d) ${\cal L}_{low} = 120 \;
\mbox{fb}^{-1}, {\cal L}_0 = 0 \; \mbox{fb}^{-1}$ and $D=$ 21 GeV. The
analysis is with the parameters of Case~I.}
\label{fig:ellipse}
\end{figure}


\begin{figure}
\noindent
\hspace*{1cm}
\fig{8cm}{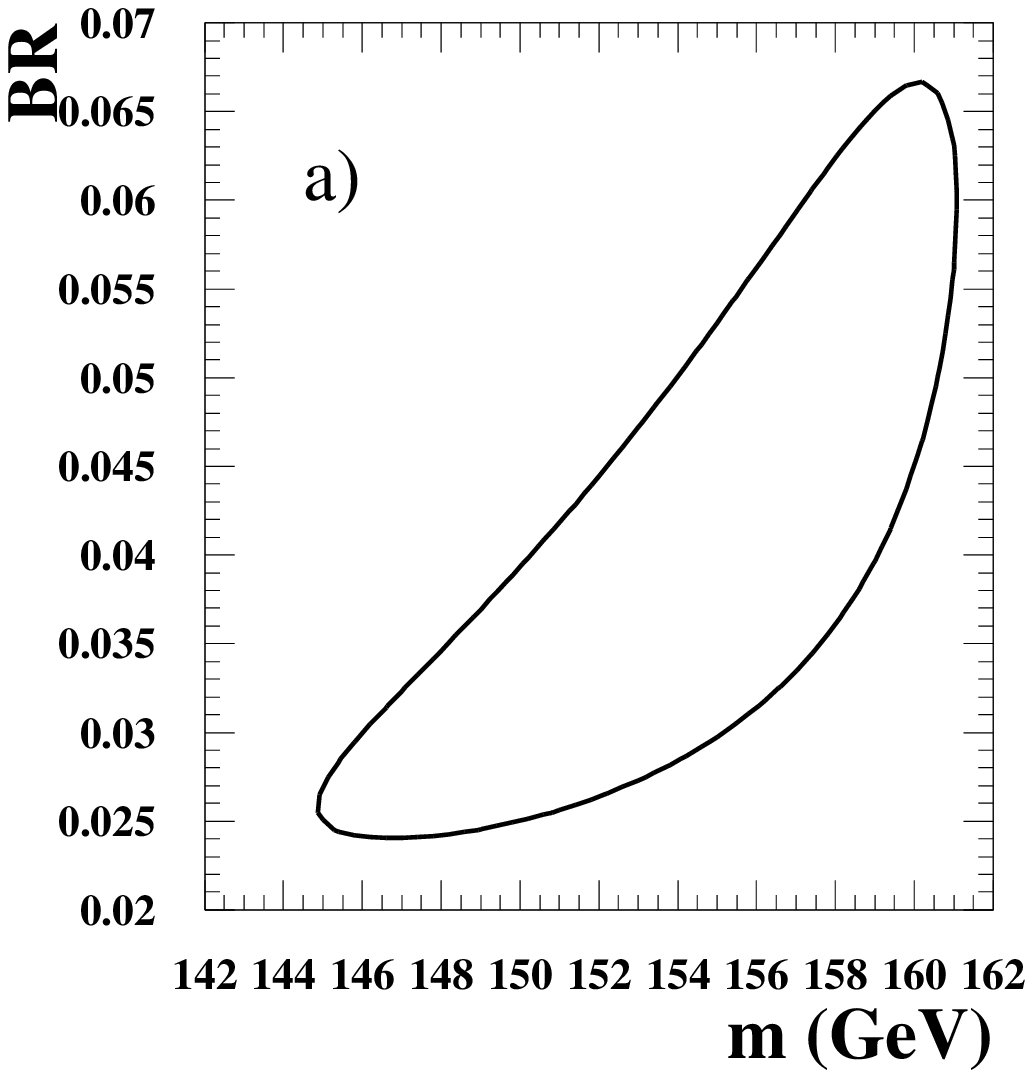} \hspace*{-1cm} \fig{8cm}{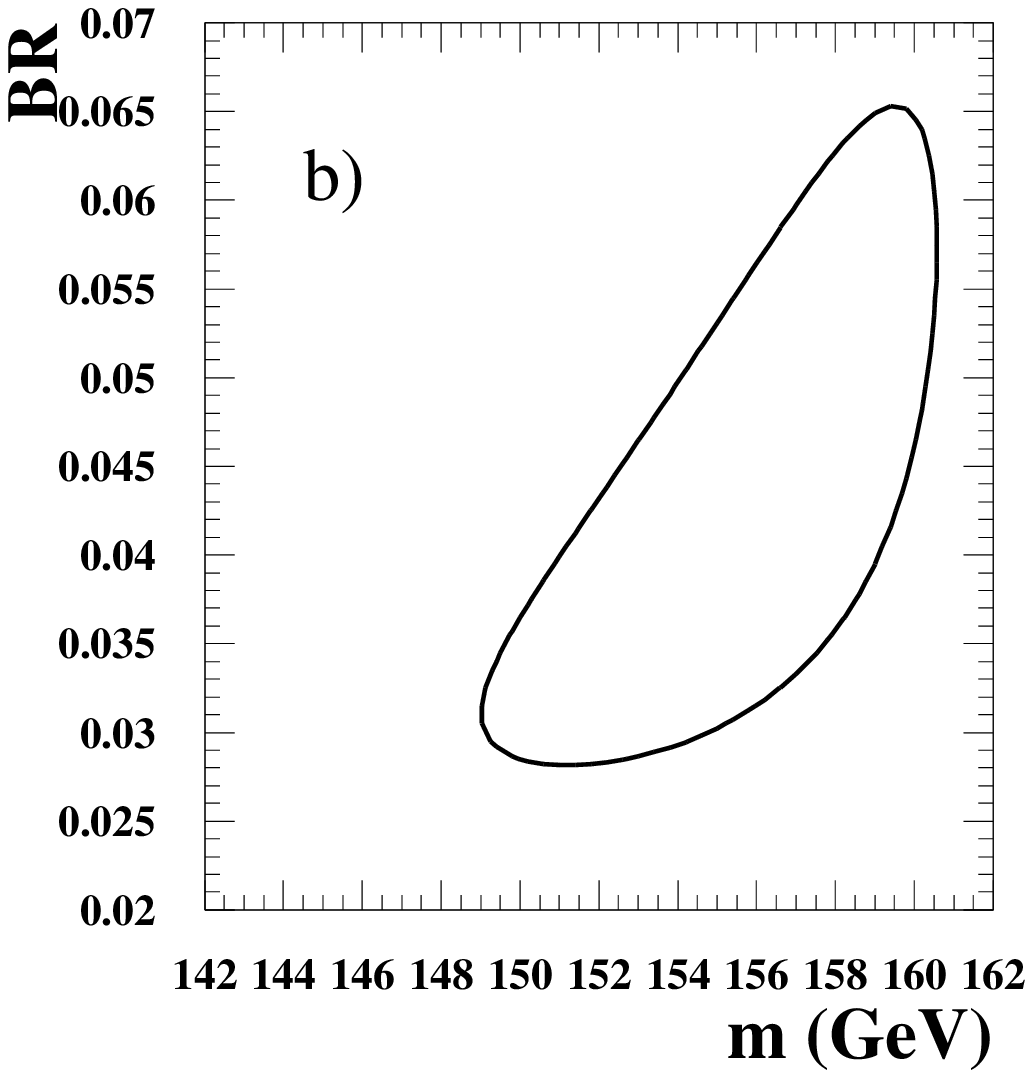}
\vspace*{-1cm}
\\
\hspace*{1cm}
\fig{8cm}{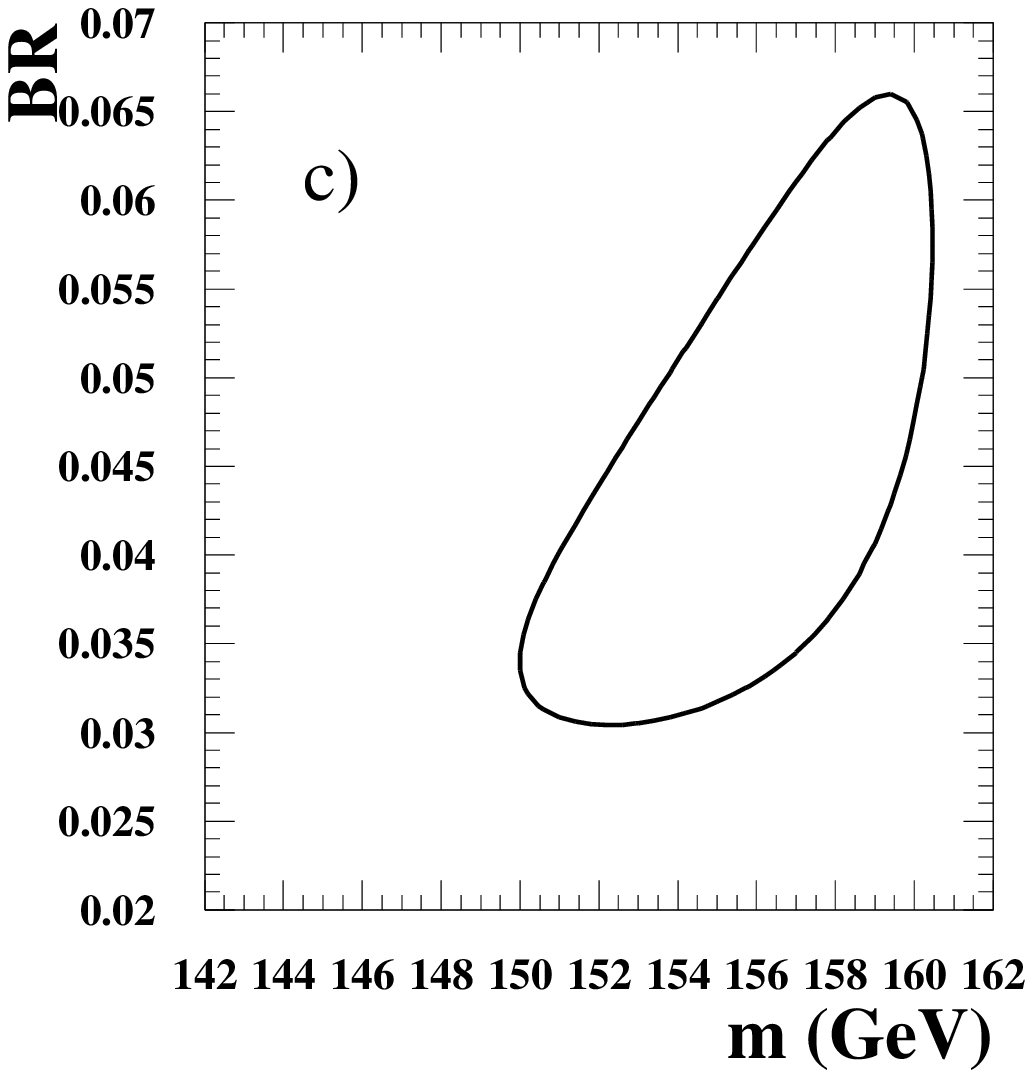} \hspace*{-1cm} \fig{8cm}{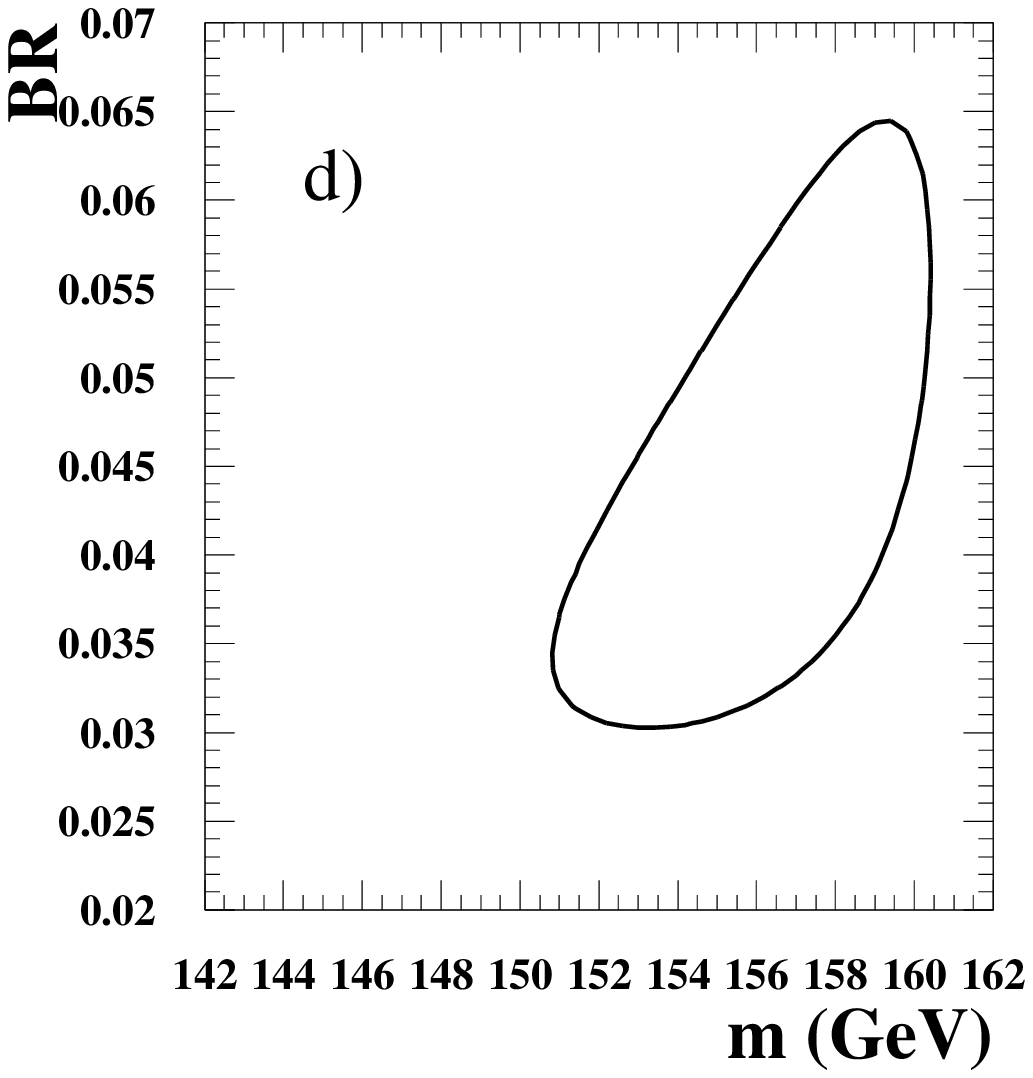}
\vspace*{0.4cm}
\caption[]{90\% CL ($\Delta\chi^2=4.6$) contours for a fixed 10 events
per point (for the input theory) at lower energies, with $N = 3, {\cal
L}_{low} = 100 \; \mbox{fb}^{-1}$ and ${\cal L}_0 = 20 \;
\mbox{fb}^{-1}$.  In a) $\Delta = 20$ GeV and $D=$ 19 GeV; in b) $\Delta
= 40$ GeV and $D=$ 16 GeV; in c) $\Delta = 60$ GeV and $D=$ 16 GeV, and
in d) $\Delta =$ 80 GeV and $D=$ 15 GeV. The analysis is with the
parameters of Case~I.}
\label{fig:delta}
\end{figure}

\newpage

\begin{figure}
\dofig{6in}{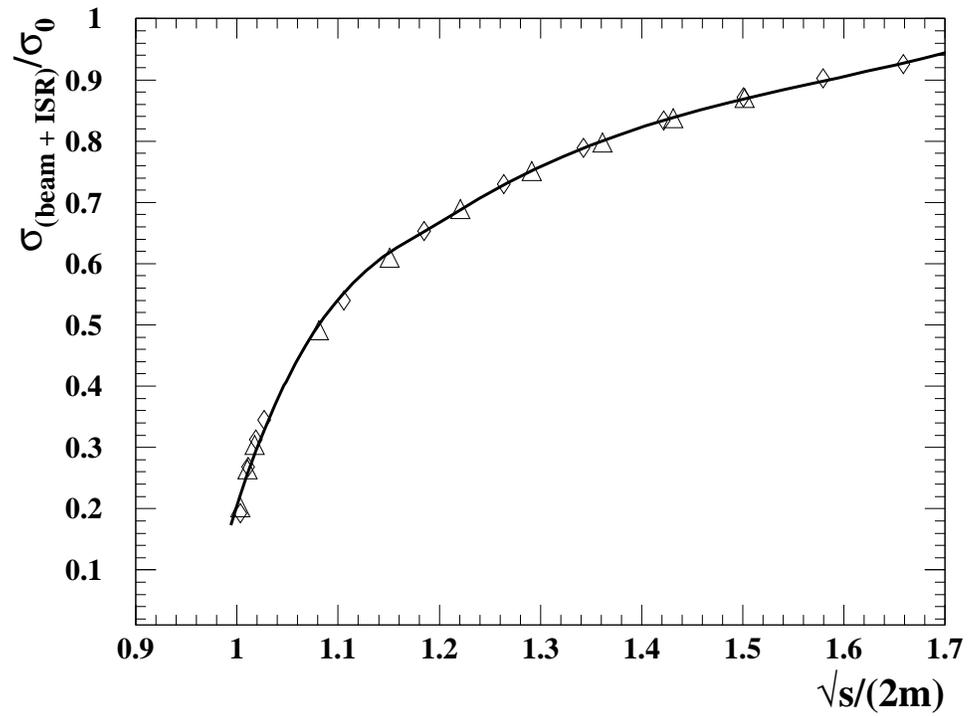}
\caption[]{Ratio of $\tnu_\tau$ (or $\tnu_{\mu}$) pair production cross
sections with and without initial state radiation and beamstrahlung
effects versus the center of mass energy divided by the sum of sneutrino
masses. Diamonds (squares) denote the results for Case I (II).}
\label{fig:beam}
\end{figure}


\begin{figure}
\dofig{6in}{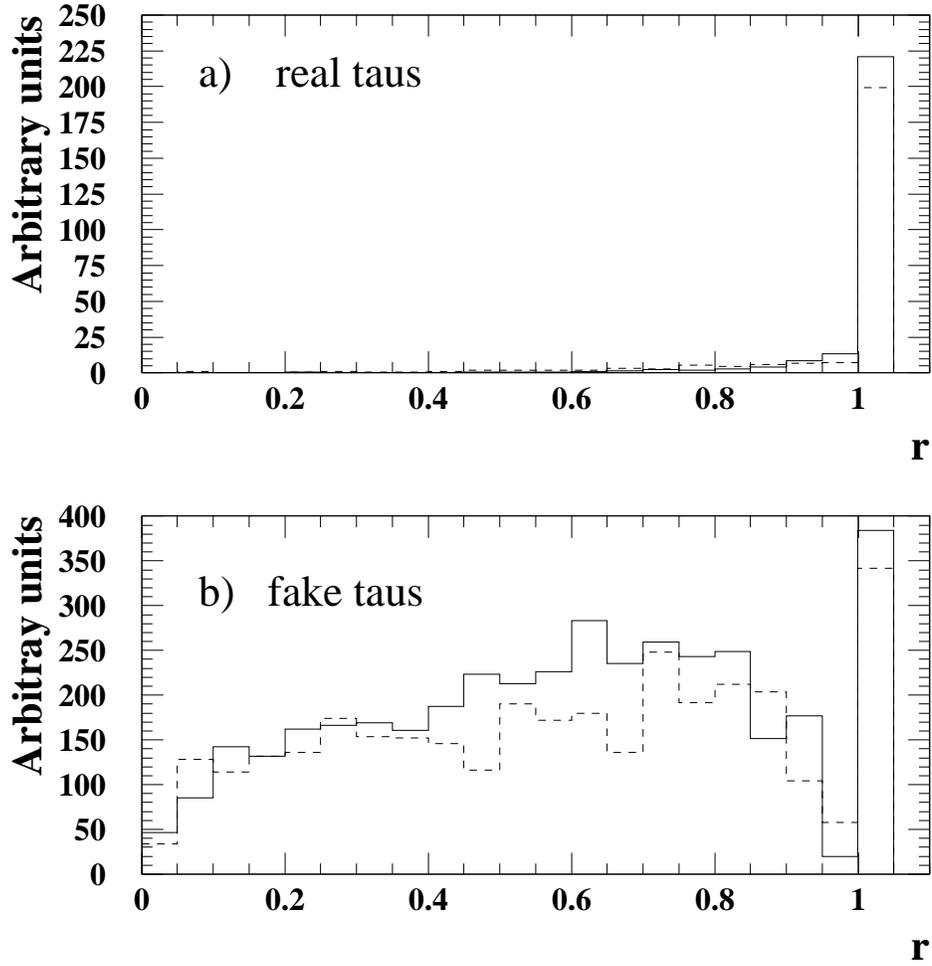}
\caption[]{The rate for {\it a})
real and {\it b})~fake tau jets versus the $\tau-c$ discrimination
parameter $r$ defined in Eq.~(\ref{eq:r}) of the text. The
dashed histogram shows the result for Case~I, while the solid histogram
shows the corresponding result for Case~II.}
\label{fig:r}
\end{figure}

\newpage

\begin{figure}
\dofig{6in}{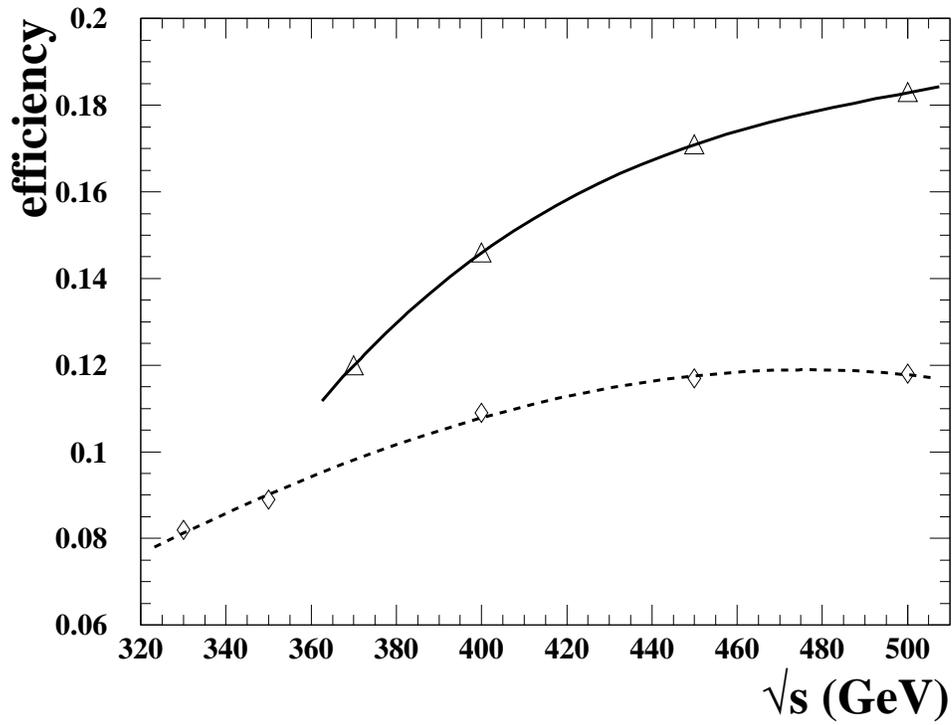}
\caption[]{Reconstruction efficiency for the $\tau \tau j j \ell
+ \eslt$ channel in Case  I (diamonds) and II (triangles). The solid and dashed
lines show the fitted efficiency functions that we use in our analysis.}
\label{fig:taueff}
\end{figure}

\newpage

\begin{figure}
\dofig{6in}{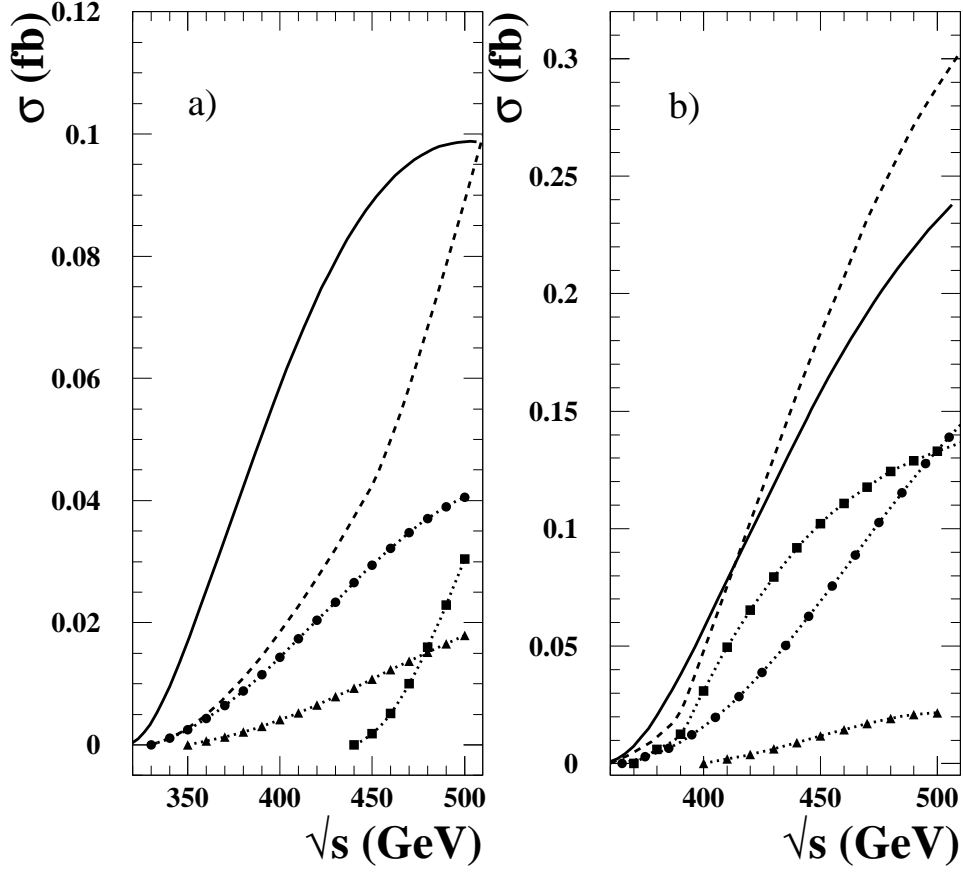}
\caption[]{Signal (solid line) and the total SUSY background (dashed
line) cross sections for $\tau \tau j j \ell + \eslt$ after including
the effects of ISR and beamstrahlung, as well as the reconstruction
efficiency, versus the center of mass energy. The circles, squares and
triangles show SUSY contamination from first two generations of charged
sleptons and sneutrinos, charginos and neutralinos, and
$\ttau_1\ttau_2+\ttau_2\ttau_2$ production, respectively. The dotted
line is the fit to these component backgrounds that we use in our
analysis. Frame {\it a}) is for Case I while frame {\it b}) is for Case
II.}
\label{fig:tauback}
\end{figure}

\newpage

\begin{figure}
\noindent
\hspace*{1cm}
\fig{8cm}{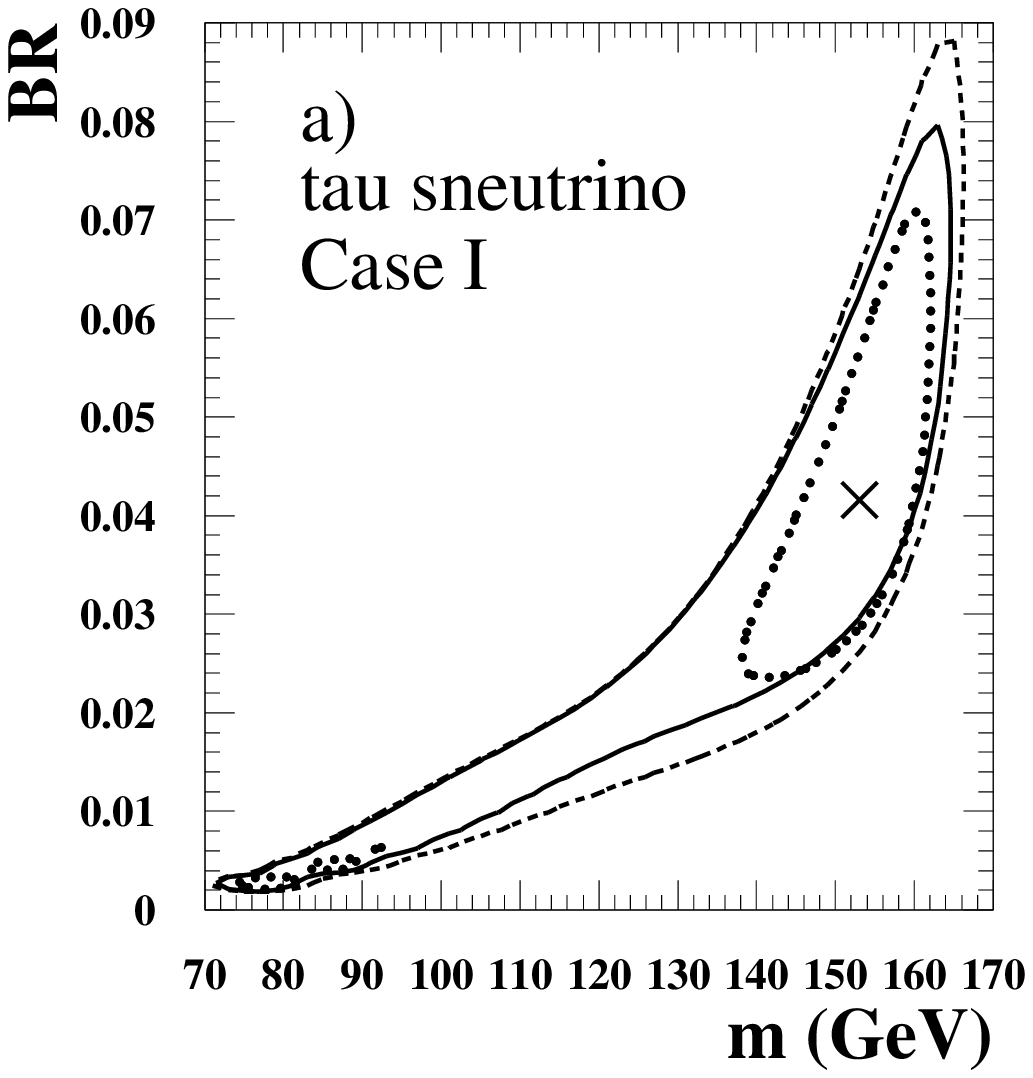} \hspace*{-1cm} \fig{8cm}{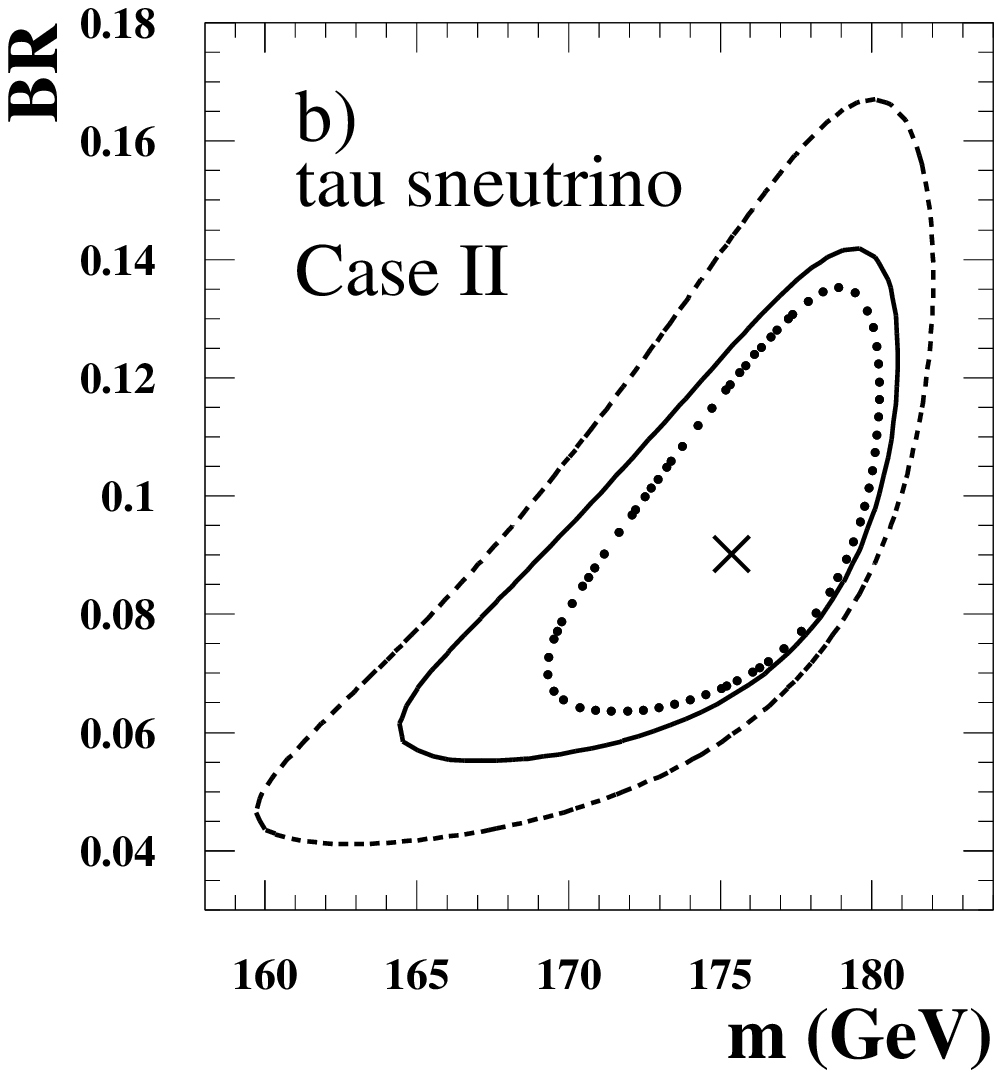}\\
\vspace*{0.4cm}
\caption[]{90\% CL ($\Delta\chi^2=4.6$) contours for a fixed 6 events
per point (for the input theory) at lower energies, with $N =$ 3, ${\cal
L}_{low} = 400 \; \mbox{fb}^{-1}, {\cal L}_0 = 100 \; \mbox{fb}^{-1}, $
and $\Delta = 60$ GeV. In both frames the solid (dashed) line stands for
limits without (with) SUSY backgrounds, while the dotted line shows how
much the error ellipse shrinks for an integrated luminosity 1.5 times
higher, assuming no background.  The cross inside the ellipses shows the
best fitted point.  Frame {\it a}) shows the result for Case~I and has
$D=$42 (30) GeV for the solid and dashed (dotted) ellipses, while frame
{\it b}) which is for Case II, has $D=$ 23 (18) GeV for the solid and
dashed (dotted) ellipse. The cross shows the best fit.}
\label{fig:tauellipse}
\end{figure}

\newpage

\begin{figure}
\dofig{6in}{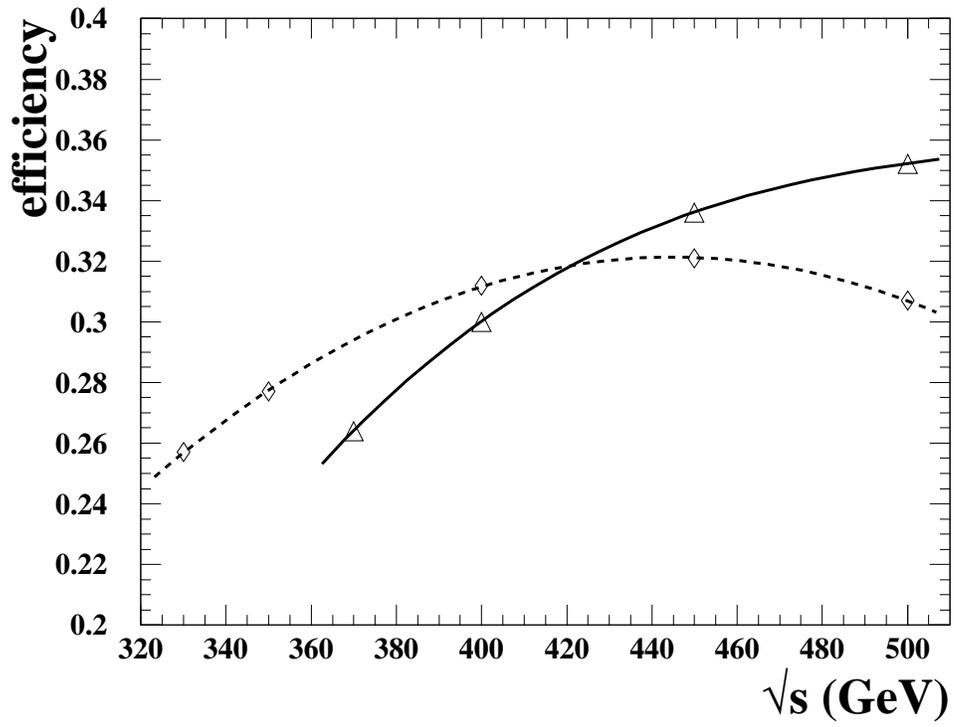}
\caption[]{Same as in Fig.~\ref{fig:taueff} but for the
$\mu \mu j j \ell + \eslt$ channel.}
\label{fig:mueff}
\end{figure}


\begin{figure}
\dofig{6in}{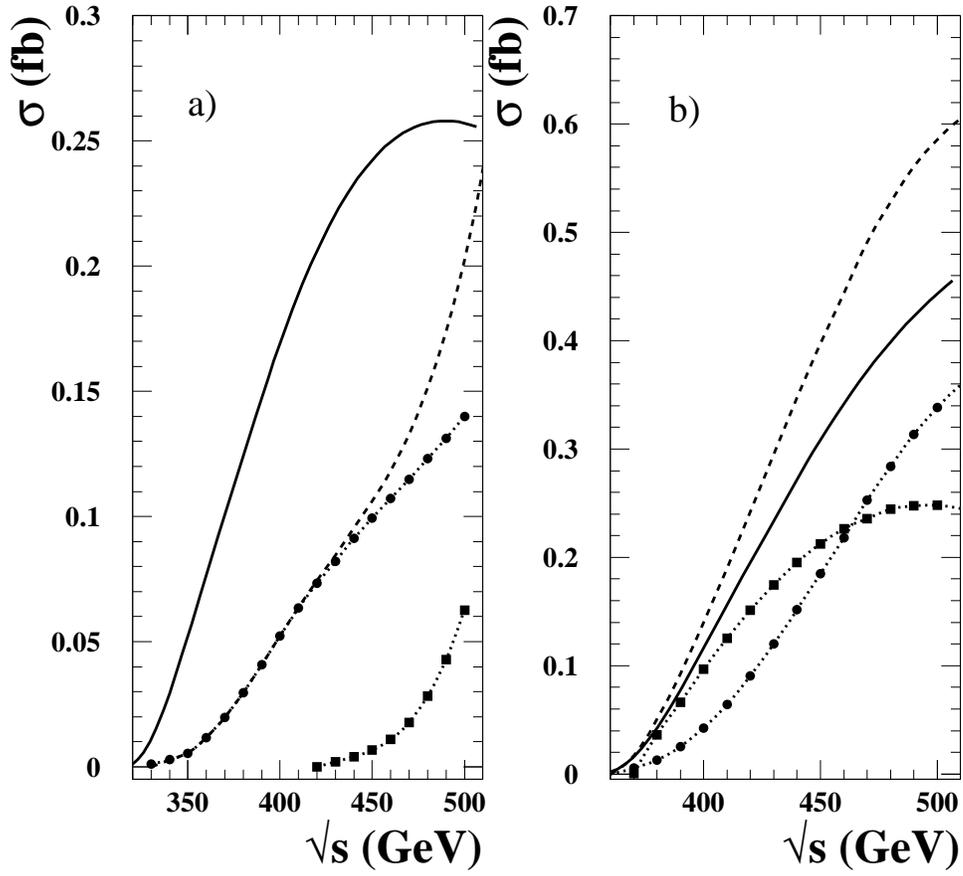}
\caption[]{Same as in Fig.~\ref{fig:tauback} for the
$\mu \mu j j \ell + \eslt$ events, except that the circles include
contamination from all charged sleptons including smuons.}
\label{fig:muback}
\end{figure}

\newpage

\begin{figure}
\noindent
\hspace*{1cm}
\fig{8cm}{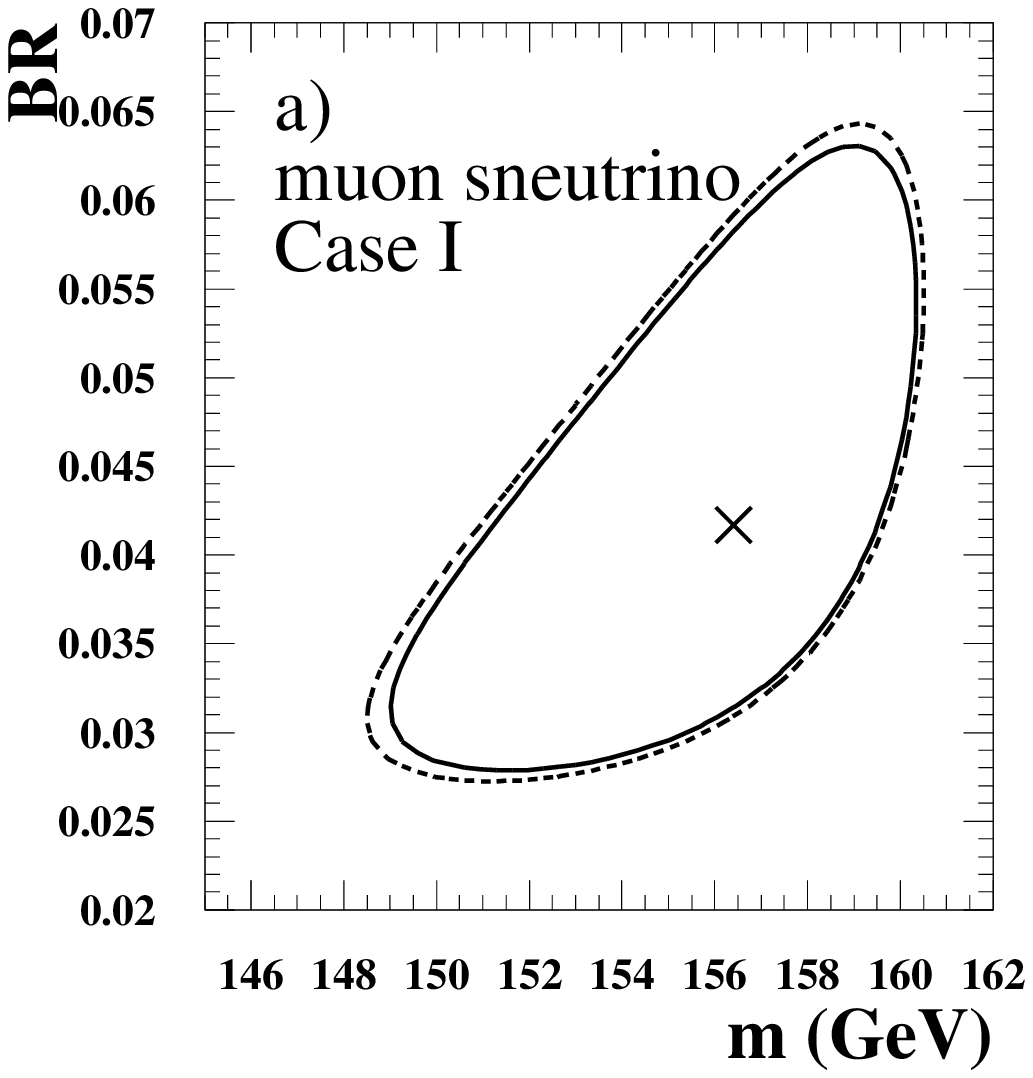} \hspace*{-1cm} \fig{8cm}{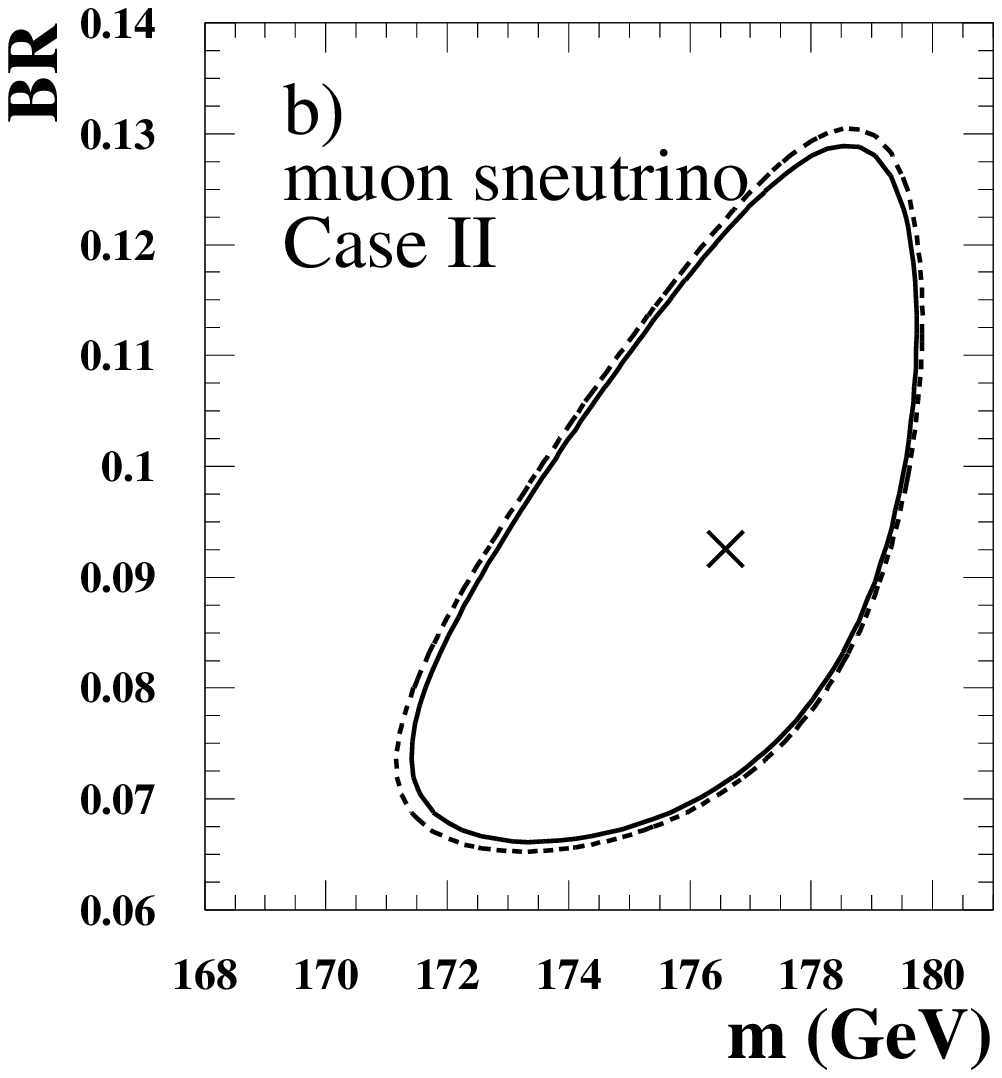}\\
\vspace*{0.4cm}
\caption[]{90\% CL ($\Delta\chi^2=4.6$) contours for a fixed 6 events
per point (for the input theory) at lower energies, with $N =$ 3, ${\cal
L}_{low} = 400 \; \mbox{fb}^{-1}, {\cal L}_0 = 100 \; \mbox{fb}^{-1}, $
and $\Delta = 60$ GeV. In both frames the solid (dashed) line stands for
limits without (with) SUSY backgrounds. In a) the result is for Case I,
with $D=$ 18 GeV, and in b) Case II, with $D=$ 14 GeV. The cross shows
the best fit. }
\label{fig:muellipse}
\end{figure}

\newpage

\begin{figure}
\noindent
\hspace*{1cm}
\fig{8cm}{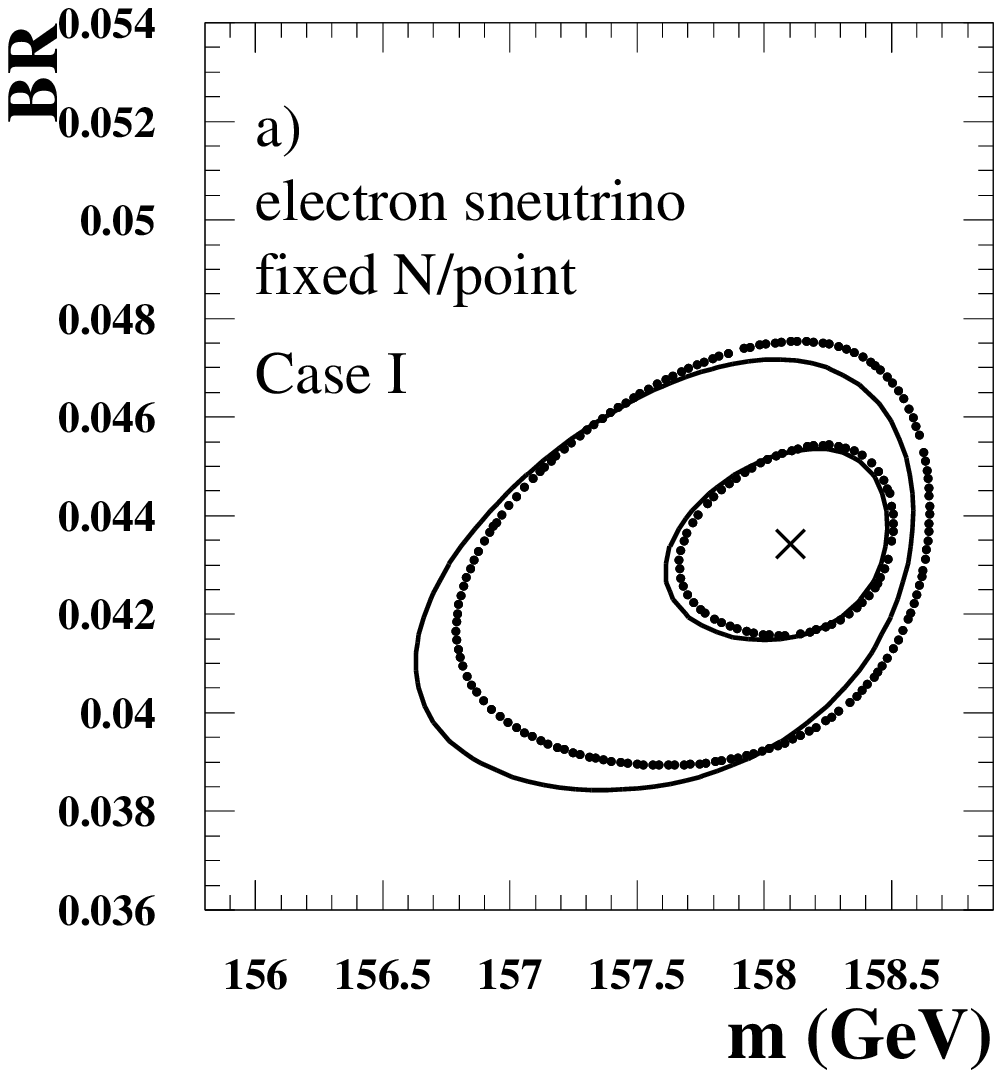} \hspace*{-1cm} \fig{8cm}{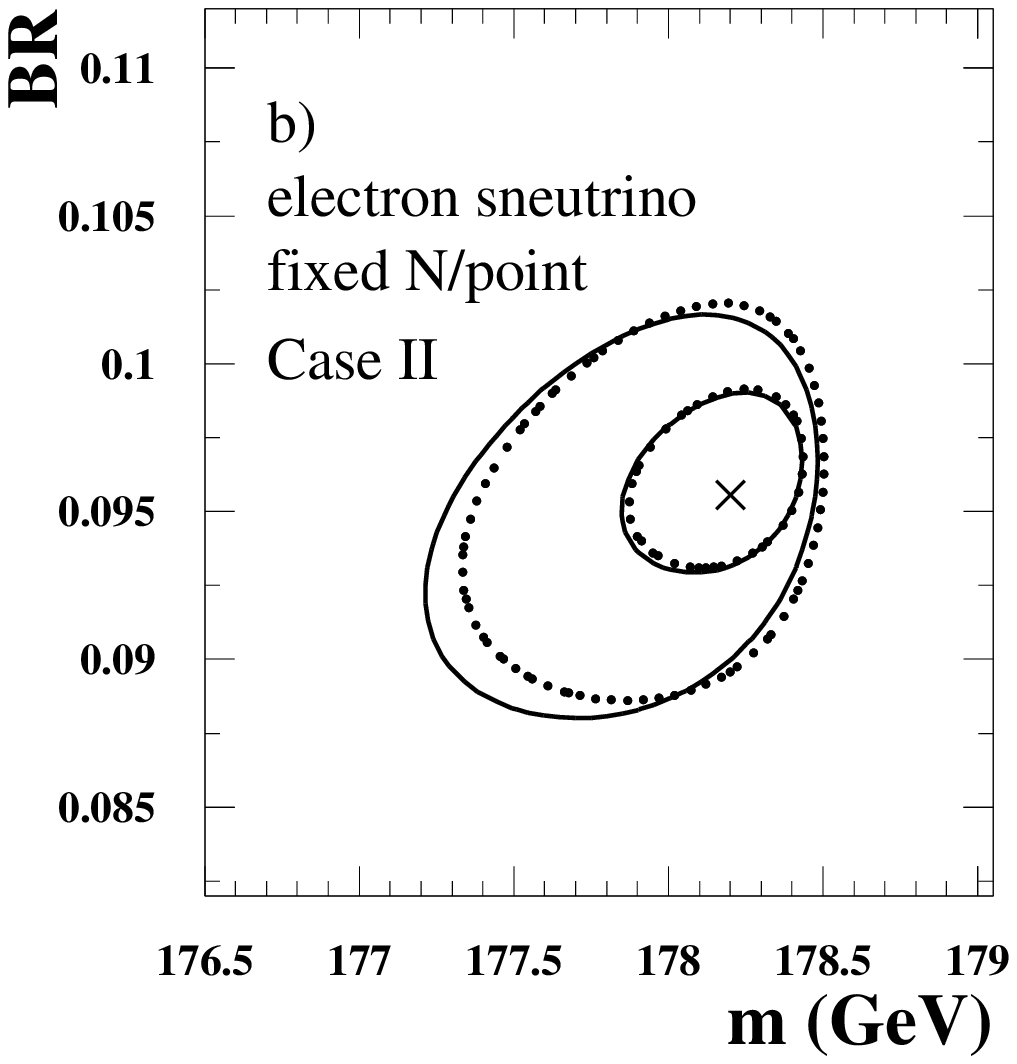}
\vspace*{-1cm}
\\
\hspace*{1cm}
\fig{8cm}{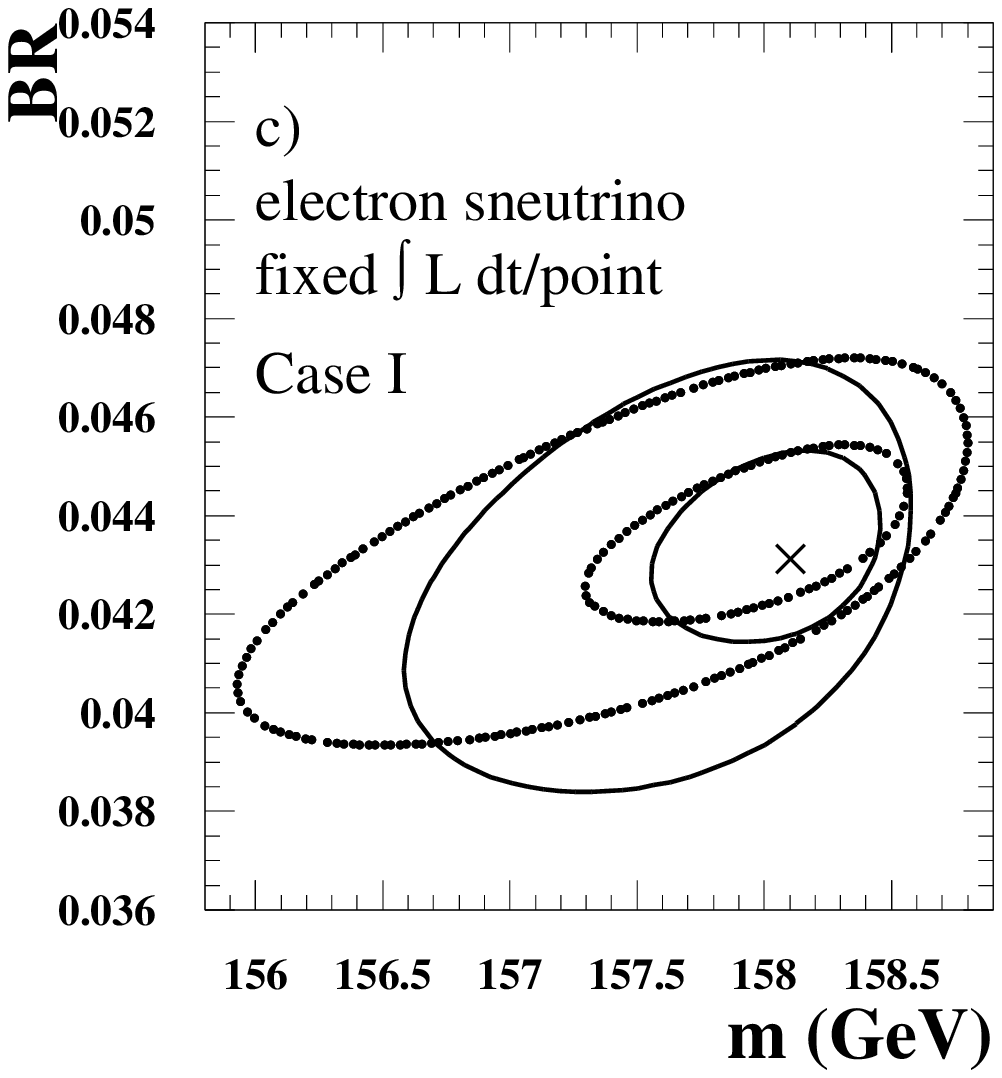} \hspace{-1cm} \fig{8cm}{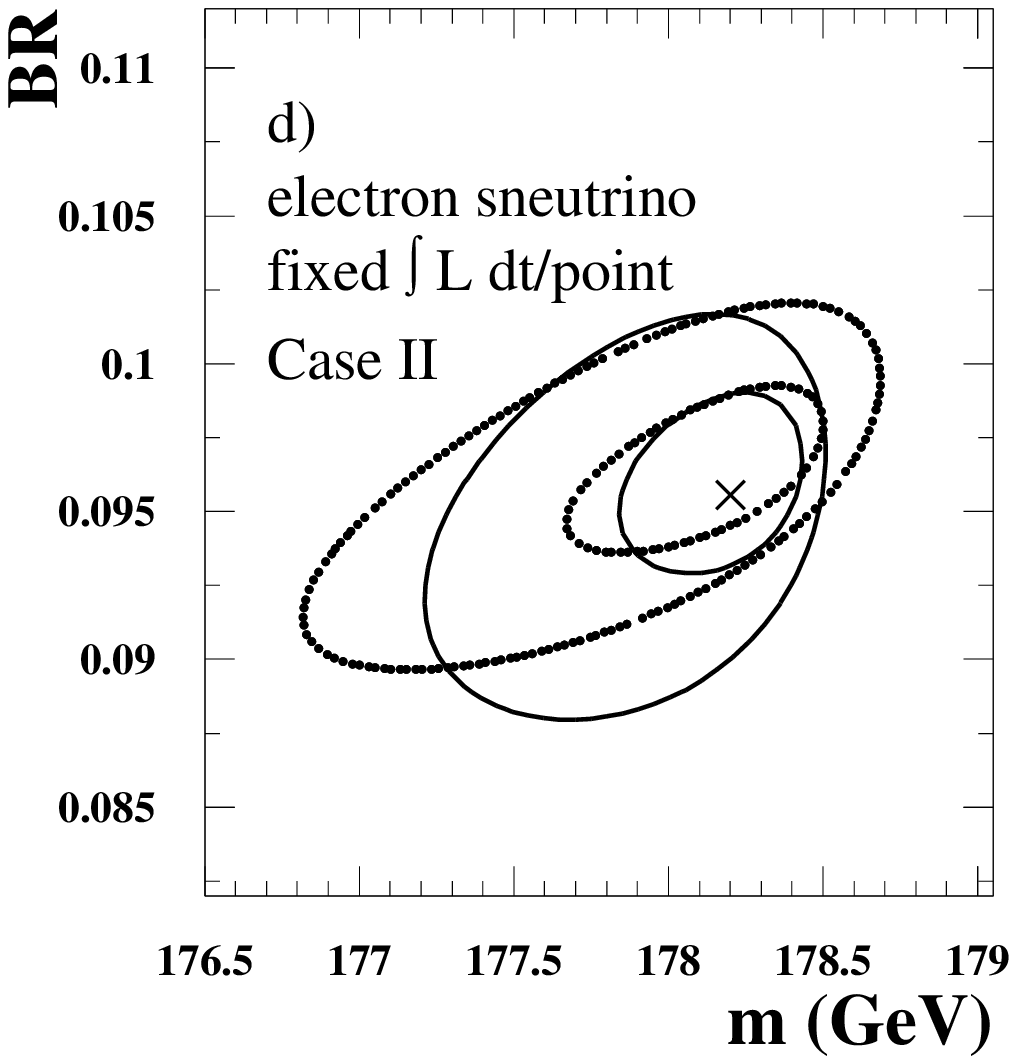}
\vspace*{0.5cm}
\caption[]{90\% CL ($\Delta\chi^2=4.6$) contours in the $m(\tnu_e)-BR$ plane, with $N=3$ and
a minimum of six signal events for each energy point after
reconstruction efficiency, beamstrahlung and ISR for Case~I (first
column) and Case~II (second column). The solid (dotted)
contours correspond to $\Delta = 1$~GeV (30~GeV). The inner (outer)
ellipses correspond to a total integrated luminosity of 500~fb$^{-1}$
(120~fb$^{-1}$) of which 100~fb$^{-1}$ (20~fb$^{-1}$) is at $\sqrt{s}=500$~GeV. In the first
row, the luminosity is distributed so that
there are an equal number of events at each of the three low energy
points, while in the second row the luminosity is equally shared between
the three points. The cross shows the best fit for the $\Delta=1$~GeV
scan with 500~fb$^{-1}$.}
\label{fig:eellipse}
\end{figure}


\begin{figure}
\noindent
\dofig{6in}{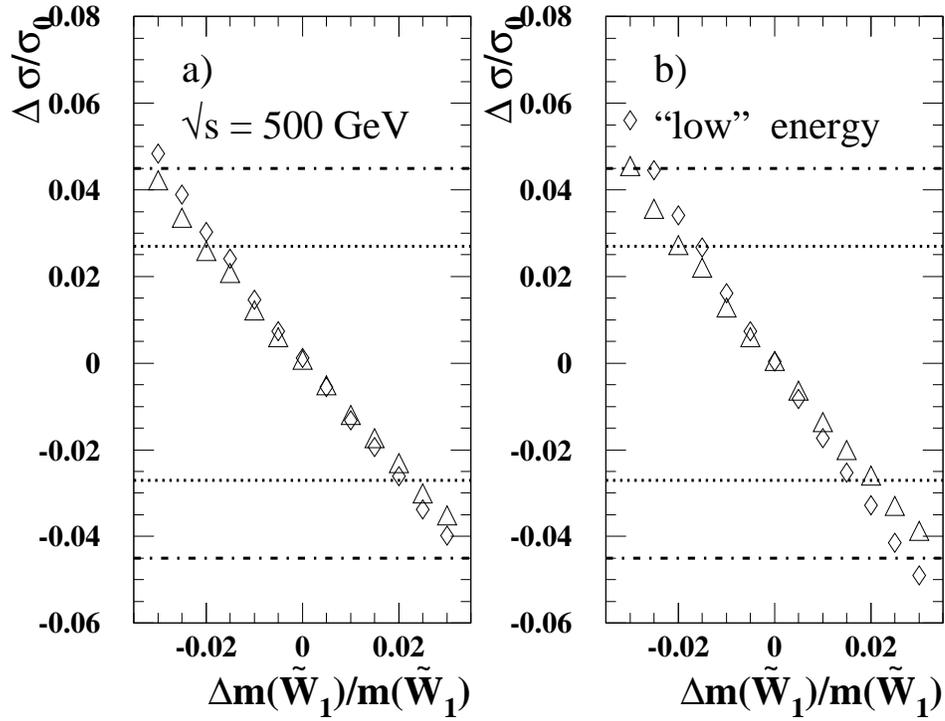}
\caption[]{The maximum fractional change in $\sigma(e^+e^- \to
\tnu_e\tnu_e$) from variation of the $\tw_1$ and $\tw_2$ masses within
their expected uncertainty as a function of the precision of $m(\tw_1)$
for {\it a})~$\sqrt{s}=500$~GeV, and {\it b})~$\sqrt{s}=2m(\tnu_e)+2$~GeV.
We assume that $\Delta m(\tw_2)/m(\tw_2) = 5\Delta
m(\tw_1)/m(\tw_1)$. The diamonds show the results for Case~I whereas the
triangles show the same for Case~II.}
\label{fig:unc}
\end{figure}

\end{document}